\documentclass[reprint,amsmath,amssymb,aps,pra]{revtex4-2}

\usepackage{amsmath}
\usepackage{graphicx}
\usepackage{bm}

\usepackage[capitalise]{cleveref}
\usepackage{makecell}
\usepackage{siunitx}

\newcommand*{\dd}{\ensuremath{\mathrm d}}
\newcommand*{\ee}{\ensuremath{\mathrm e}}
\newcommand*{\ii}{\ensuremath{\mathrm i}}

\renewcommand*{\phi}{\varphi}
\DeclareMathOperator\erfc{erfc}
\DeclareMathOperator\sgn{sgn}

\begin{document}

\title{Lattice Sums Accommodating Multiple Sublattices for Solutions of the Helmholtz Equation in Two and Three Dimensions}

\author{Dominik Beutel}
\email{dominik.beutel@kit.edu}
\affiliation{Institute of Theoretical Solid State Physics, Karlsruhe Institute of Technology (KIT), 76131 Karlsruhe, Germany}
\date{\today}
\author{Ivan Fernandez-Corbaton}
\affiliation{Institute of Nanotechnology, Karlsruhe Institute of Technology (KIT), 76344 Eggenstein-Leopoldshafen, Germany}
\author{Carsten Rockstuhl}
\affiliation{Institute of Theoretical Solid State Physics, Karlsruhe Institute of Technology (KIT), 76131 Karlsruhe, Germany}
\affiliation{Institute of Nanotechnology, Karlsruhe Institute of Technology (KIT), 76344 Eggenstein-Leopoldshafen, Germany}

\begin{abstract}
The evaluation of the interaction between objects arranged on a lattice requires the computation of lattice sums. A scenario frequently encountered are systems governed by the Helmholtz equation in the context of electromagnetic scattering in an array of particles forming a metamaterial, a metasurface, or a photonic crystal. 
While the convergence of direct lattice sums for such translation coefficients is notoriously slow, the application of Ewald's method converts the direct sums into exponentially convergent series. We present a derivation of such series for the 2D and 3D solutions of the Helmholtz equation, namely spherical and cylindrical solutions. When compared to prior research, our novel expressions are especially aimed at computing the lattice sums for several interacting sublattices in 1D lattices (chains), 2D lattices (gratings), and 3D lattices.  We verify our results by comparison with the direct computation
of the lattice sums.

\end{abstract}

\maketitle

The calculation of lattice sums for solutions of the Helmholtz equation appears regularly in various fields of physics, such as electrodynamics, solid-state physics, or acoustics~\cite{varadan1980,waterman2009}. A particularly useful tool to treat those sums is Ewald's method~\cite{ewald1921} with it various applications \cite{babicheva2021,berkhout2020,chen2017,cummins1976,gallinet2010,goodarzi2021,hu2021,lovat2008,lunnemann2013,lunnemann2016,rahimzadegan2022,rider2022,stefanou1998,stefanou2000,yermakov2018}. There, the slowly converging series is split into two parts. One of these parts  converges rapidly in real space, and the other one converges rapidly after a transition into reciprocal space.
In-depth discussions on this method applied to the Helmholtz equations can be found in several reviews~\cite{moroz2006,linton2010}, and there exist various derivations for special cases~\cite{capolino2005,capolino2007,chin1994,craeye2006,belov2006,dienstfrey2001,jandieri2019,mcphedran2000,moroz2002}. 

\begin{figure*}
    \centering
    \includegraphics[width=\linewidth]{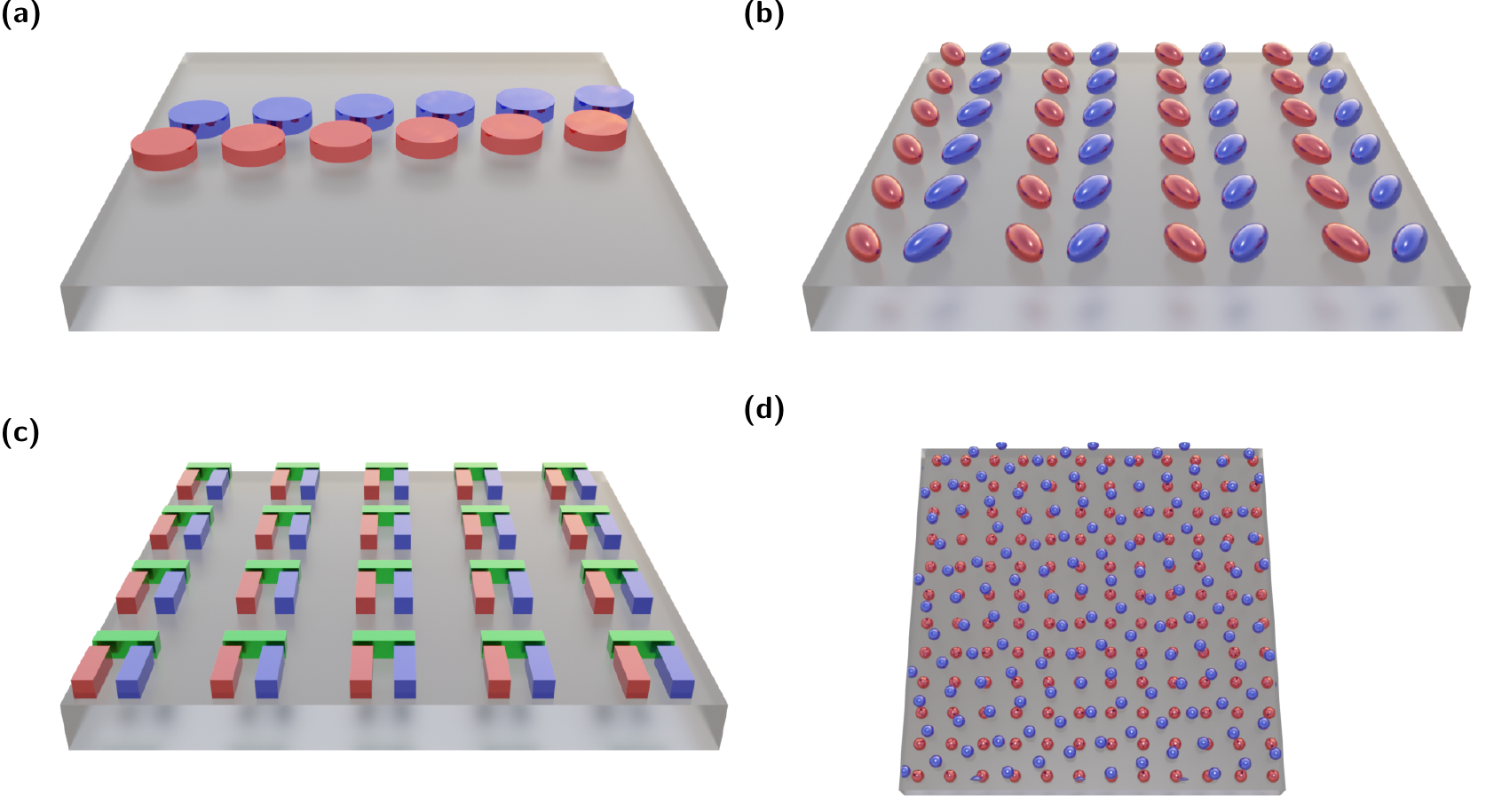}
    \caption{
        Illustrations for contemporary photonic materials featuring multiple sublattices. From panels (a) to (d) these are zigzag chains that mimick the Su-Schrieffer-Heeger model, metasurfaces where bound states in continuum can be observed, plasmonically induced transparency in dolmen structures, and Moir\'{e} lattices.
    }\label{fig:examples}
\end{figure*}

However, what has not yet been fully considered are lattices with multiple sublattices. Typical systems with multiple sublattices in different dimensions are chains with alternating distances between particles~\cite{su2013}, zigzag chains~\cite{kruk2017}, helical structures~\cite{chen2010}, or structures based on the honeycomb lattice~\cite{rechtsman2013}. Also, having many particles in one unit cell may require such lattice sums~\cite{theobald2021}. Photonic materials with multiple sublattices emerge in the context of many contemporary photonic materials. Examples are Su-Schrieffer-Heeger chains found in topological photonics~\cite{moritake2022}, structures with an asymmetry in their unit cell to support bound states in the continuum~\cite{koshelev2018}, dolmen structures to observe plasmonically induced transparency~\cite{sun2017}, or Moir\'{e} lattices~\cite{wang2020}. Conceptional illustrations showing such contemporary artificial photonic materials are presented in \cref{fig:examples}. We highlight the multiple sublattices by using red, blue, and green colors for the associated particles.
To efficiently describe the optical response from such photonic materials, we develop here novel approaches for Ewald's summations for lattices that contain multiple sublattices with arbitrary relative positions with respect to each other. That holds particularly for sublattices that have a relative displacement perpendicular to the lattice.
Our sums are also useful to compute, e.g., the electromagnetic field at an arbitrary position within the unit cell, because this essentially is equivalent to a relative shift with respect to the objects in the lattice~\cite{necada2021}.

The goal of this article is two-fold: first, we present a way of deriving exponentially convergent expressions by extending an existing approach~\cite{solbrig1982} that, second, is applicable to many of these problem of arbitrary positions in the unit cell of the lattice. Our novel approach reproduces known results for 2D lattices and spherical wave solutions~\cite{kambe1967,kambe1968}, but it is also able to derive expressions for multiple sublattices in the 2D case of cylindrical wave solutions and for 1D lattices.

The outline of the article is as follows. In Sec.~\ref{sec:problem}, we give a definition of the lattice sum and the notation used. In Sec.~\ref{sec:real}, we treat the real space sum and, in Sec.~\ref{sec:reciprocal}, the reciprocal space sum. For the reciprocal space sum, we derive closed form expressions for all cases individually. We conclude with a comparison of the presented expressions with results obtained from the direct summation approach in Sec.~\ref{sec:comparison} and an example application of the lattice sums to the T-matrix method~\cite{waterman1965,mishchenko2020} in Sec.~\ref{sec:example}. These numerical examples clearly demonstrate the usefulness and forte of our novel expressions. 

The main results needed for an implementation are \cref{eq:realsum:cw,eq:realsum:sw} for the real space sum of cylindrical and spherical solutions, respectively, one of the \cref{eq:rec:fulllattice,eq:rec:sw2d,eq:rec:sw1d,eq:rec:cw1d} for the reciprocal space sum depending on the spatial dimension and the lattice dimension, and \cref{eq:zero} as correction term for the origin contribution.

\section{Problem statement and notation}\label{sec:problem}
We define the lattice sum as
\begin{equation}\label{eq:prob}
D_{d, \nu}(\Lambda_{d'}, k, \bm k_\parallel, \bm r)
= \sideset{}{'}{\sum}_{\bm R \in \Lambda_{d'}}
f_{d, \nu}(k, -\bm r - \bm R)
\ee^{\ii \bm k_\parallel \bm R}
\end{equation}
and derive expressions for the spatial dimensions $d \in \{2, 3\}$. The second index $\nu$ is a placeholder for the parameters of the function $f_{d, \nu}$. 
The lattice $\Lambda_{d'}$ is a set containing the $d' \leq d$ dimensional lattice vectors defined by
\begin{equation}\label{eq:lattice}
    \Lambda_{d'} = \left\{\sum_{i=1}^{d'} n_i \bm a_i ~|~ n_i \in \mathbb{Z} \right\}
    \,,
\end{equation}
where $\bm a_i$ are the basis vectors of the lattice. We use $k$ for the wave number and $\bm k_\parallel$ for the wave vector components in the $d'$ dimensional reciprocal space. Later, we use the notation $\Lambda_{d'}^\ast$ for the reciprocal space lattice defined analogously to \cref{eq:lattice} with basis vectors $\bm b_j$ satisfying $\bm a_i \bm b_j = 2\pi\delta_{ij}$.
The vector $\bm r \in \mathbb R^d$ describes the shift between sublattices, and it can be decomposed into a tangential component $\bm r_\parallel \in \mathbb R^{d'}$ and a normal component $\bm r_\perp \in \mathbb R^{d-d'}$ with respect to the vectors of the lattice $\Lambda_{d'}$.

On the right hand side of \cref{eq:prob}, the sum includes all lattice points with the exception that in the case of $\bm r + \bm R = 0$, i.e., if $\bm r$ coincides with a lattice point we omit that specific contribution. We use the prime next to the summation sign as a reminder of this ommission. Each term of the sum contains a phase factor and the scalar solutions of the Helmholtz equation for the chosen dimension $d$, namely
\begin{equation}
f_{2, \nu}(k, \bm r) = H^{(1)}_{m}(k |\bm r|) \ee^{\ii m \phi_{\bm r}}
\end{equation}
and
\begin{equation}
f_{3, \nu}(k, \bm r) = h^{(1)}_{l}(k |\bm r|) Y_{lm}(\theta_{\bm r}, \phi_{\bm r})\,.
\end{equation}
Thus, the index $\nu$ stands for $m \in \mathbb Z$, if $d = 2$, and for $l \in \mathbb N_0$ and $m \in \{-l, -l+1, \cdots, l\}$, if $d = 3$.
The functions $H_m^{(1)}(x)$ are the Hankel functions of the first kind, $h_l^{(1)}(x)$ are the spherical Hankel functions of the first kind, and $Y_{lm}(\theta, \phi)$ are the spherical harmonics. See appendix~\ref{app:sphharm} for the used normalization convention.
We also use the notation $Y_{l m}(\bm r) = Y_{l m}(\theta_{\bm r}, \phi_{\bm r})$, where $\cos\theta_{\bm{r}} = \tfrac{z}{|\bm r|}$ and $\tan\phi_{\bm{r}} = \tfrac{y}{x}$ are the polar and azimuthal angle of the vector $\bm{r}$.

\begin{figure*}
    \centering
    \includegraphics[width=\linewidth]{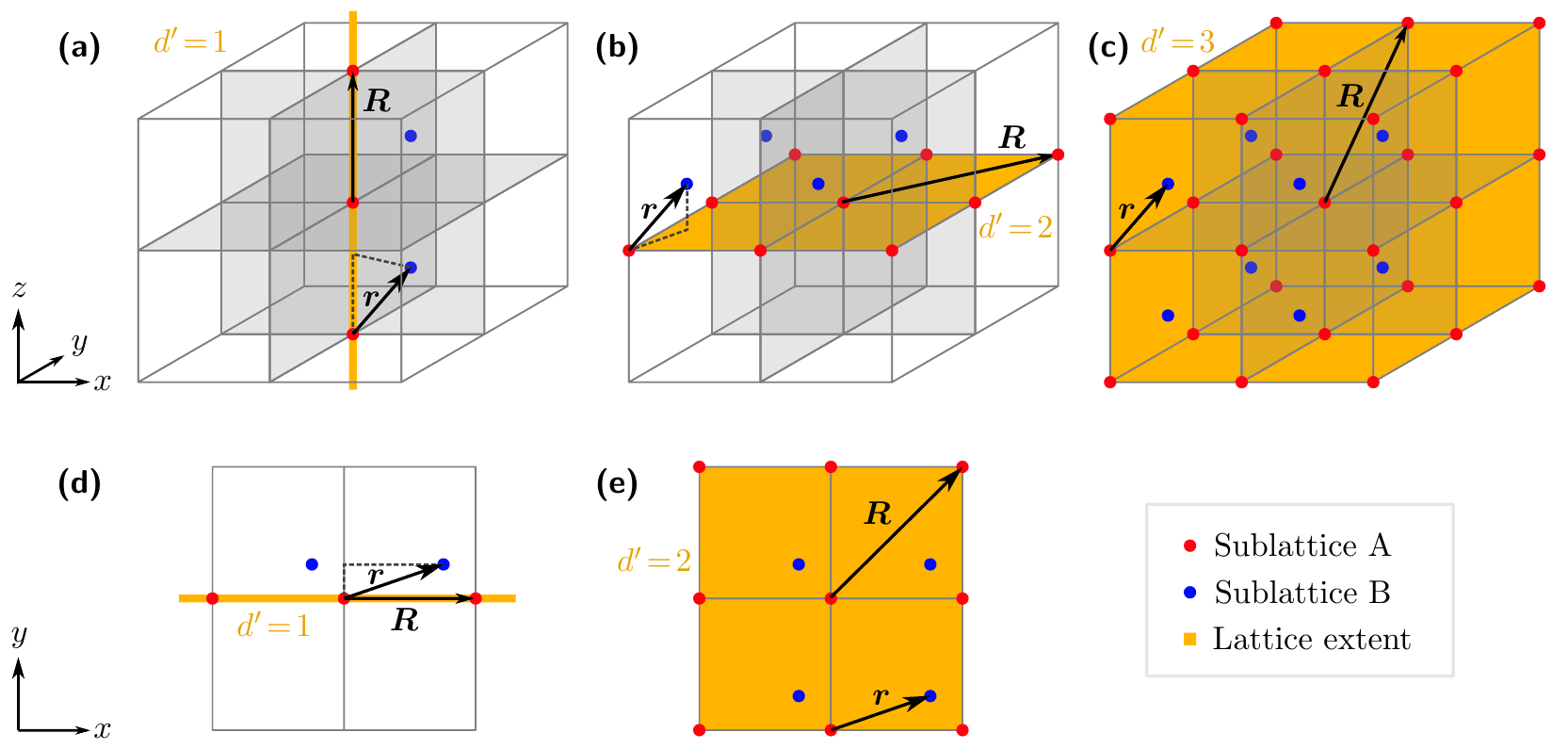}
    \caption{
        Layout of the geometry in the lattice summations. In all cases, we show two different sublattices in red and blue.  The shift between these sublattices is described by $\bm r$. The dotted lines indicate the decomposition of that shift into $\bm r_\perp$ and $\bm r_\parallel$. In panels (a), (b), and (d), the different sublattices are not required to be in the same plane or line. The vector $\bm R$ is a lattice vector. The first row with panels (a) to (c) shows the case of spatial dimension $d = 3$ and $d' \in \{1,2,3\}$, respectively. In all three cases we use the coordinate system indicated on the left. The second row with pannels (d) and (e) shows the cases for $d = 2$, again, with coordinates as shown on the left. The orange color shows the spatial domain of the lattice summation. The parallel component of the wave vector $\bm k_\parallel$ must lie in this domain.  
    }\label{fig:layout}
\end{figure*}

These definitions lead to five different possible cases shown in \cref{fig:layout}. For $d = 3$, the lattice can have $d' \in \{1, 2, 3\}$ as shown in panels (a) to (c). For $d = 2$ the lattice can have $d' \in \{1, 2\}$.

The starting point for the evaluation of the sum expressed in \cref{eq:prob} are the representations~\cite{eyert2012}
\begin{equation}
    H_{l}^{(1)}(x)
    = \frac{(-1)^\frac{l - |l|}{2} 2}{\ii \pi}
    x^{|l|}
    \int\limits_{(0)}^\infty \dd t\,t^{2|l| - 1}
    \ee^{-\frac{x^2t^2}{2} + \frac{1}{2t^2}}
    ,~l \in \mathbb{Z}
\end{equation}
and
\begin{equation}
    h_{l}^{(1)}(x)
    = -\ii \sqrt{\frac{2}{\pi}}
    x^l
    \int\limits_{(0)}^\infty \dd t\,t^{2l}
    \ee^{-\frac{x^2t^2}{2} + \frac{1}{2t^2}}
    ,~l \in \mathbb{N}_0
    \,,
\end{equation}
where we exchanged the azimuthal order index $m$ for the case $d = 2$ with the letter $l$ to highlight the similarity between the expressions for both cases. 
To ensure convergence, the integration contour has to be chosen such that $|\Im (t)| > |\Re (t)|$ for $t \rightarrow 0$  at the lower boundary. The brackets around the lower boundary's value are used as a reminder for that choice of integration contour.
Inserting these representations into the expression for $D_{d, \nu}$ in \cref{eq:prob}, we arrive at
    \begin{equation}\label{eq:fullexpr}
        \begin{split}
        D_{d, \nu}(\Lambda_{d'}, k, \bm k_\parallel, \bm r)
        = \sideset{}{'}{\sum}_{\bm R \in \Lambda_{d'}}
        \frac{2 \ee^{\ii \bm k_\parallel \bm R}}{\ii \pi}
        (k|\bm r + \bm R|)^{|l|}&
        \\
        \cdot
        \int\limits_{(0)}^\infty \dd t\,t^{2|l| - 3 + d}
        \ee^{-\frac{(k|\bm r + \bm R|t)^2}{2} + \frac{1}{2t^2}}&
        \\
        \cdot
        \begin{cases}
            (-1)^\frac{l - |l|}{2} \ee^{\ii l \phi_{-\bm r - \bm R}} & d = 2\\
            \sqrt{\frac{\pi}{2}} Y_{lm}(-\bm r - \bm R) & d = 3
        \end{cases}&
        \,.
        \end{split}
    \end{equation}
Now, the integration can be separated at some value $\eta$, which divides the sum into a long range ($t > \eta$) and a short range ($t < \eta$) contribution, each of which can be solved separately. Indeed, this separation converts the short range contribution into a quickly convergent series in real space and the long range contribution into a quickly convergent series after a transformation into reciprocal space. The following two sections are dedicated to these spaces individually.

\section{Real space sum}\label{sec:real}
The short range part can be readily summed in real space. The only change to the expression in \cref{eq:fullexpr} is that the required integration changes to
\begin{equation}\label{eq:int:real}
    I_n(x, \eta)
    = \int\limits_{\eta}^\infty \dd t\,t^{n}
    \ee^{-\frac{x^2t^2}{2} + \frac{1}{2t^2}}
    \,,
\end{equation}
where $n \geq -1$.
For a numerical implementation, this integral can be evaluated by recurrence (see appendix~\ref{app:realrecint}). 
However, the lattice sum is evaluated in general with the expressions
\begin{equation}\label{eq:realsum:cw}
    \begin{split}
        D_{2,l}^{(2)}(\Lambda_{d'}, k, \bm k_\parallel, \bm r)
        = \frac{(-1)^\frac{l - |l|}{2} 2}{\ii \pi}
        \sideset{}{'}{\sum}_{\bm R \in \Lambda_{d'}}
        \Big(
        (k|\bm r + \bm R|)^{|l|}
        \\
        \cdot
        I_{2|l| - 1}(k |\bm r + \bm R|, \eta)
        \ee^{\ii l \phi_{-\bm r - \bm R} + \ii \bm k_\parallel \bm R}
        \Big)
    \end{split}
\end{equation}
and
\begin{equation}\label{eq:realsum:sw}
    \begin{split}
    D_{3,l m}^{(2)}(\Lambda_{d'}, k, \bm k_\parallel, \bm r)
    = -\ii \sqrt{\frac{2}{\pi}}
    \sideset{}{'}{\sum}_{\bm R \in \Lambda_{d'}}
    \Big(
    \ee^{\ii \bm k_\parallel \bm R}
    (k|\bm r + \bm R|)^{|l|}&
    \\
    \cdot
    I_{2|l|}(k |\bm r + \bm R|, \eta)
    Y_{l m}(-\bm r - \bm R)
    \Big)&\,,
    \end{split}
\end{equation}
where the total sum of \cref{eq:fullexpr} has been conventionally written in three terms as
\begin{equation}
    D_{d,\nu} = D_{d,\nu}^{(0)} + D_{d,\nu}^{(1)} + D_{d,\nu}^{(2)}\,.
\end{equation}
Here, we omitted the arguments of the different sums.
The first two terms are related to the reciprocal space sum discussed in the next chapter.
For an increasing length of the lattice vectors $\bm R$, the summands quickly decrease due to the exponential factor in \cref{eq:int:real}. Therefore, the lattice series can be truncated after including few lattice vectors.
The expressions here make no assumptions on the orientation of the lattice for the cases when $d' < d$, but we will require the specific orientations shown in \cref{fig:layout} for the reciprocal lattice sum. In the special case of $\bm r_\perp = 0$, the symmetry of the solution sets together with the orientation of the lattices can lead to simplifications (appendix~\ref{app:simplify}) that can be used to reproduce the results for this special case~\cite{linton2010}.

\section{Reciprocal space sum}\label{sec:reciprocal}

The long range contribution is summed in reciprocal space.
For the transformation into reciprocal space, a fully periodic lattice is necessary, so the inclusion of the potentially missing summand at $\bm r + \bm R = 0$ in \cref{eq:fullexpr} needs to be compensated.
This compensation contribution is independent of the lattice dimension and can be written as
\begin{equation}
    D_{d\nu}^{(0)}(\Lambda_{d'}, k, \bm k_\parallel, \bm r)
    = -\frac{\delta_{l0}\delta_{\bm r 0}}{i\pi}
    \int\limits_{(0)}^\eta \dd t\,
    \ee^{\frac{1}{2t^2}}
    \begin{cases}
        \frac{2}{t} & d = 2\\
        \frac{1}{\sqrt{2}} & d = 3
    \end{cases}
    \,,
\end{equation}
where we assume that $\bm r$ is in the Wigner-Seitz cell of the lattice, such that $\bm r + \bm R = 0$ implies $\bm r = 0 = \bm R$, and we use that all contributions except for $l = 0$ vanish due to the factor $(k |\bm r + \bm R|)^{|l|}$ in \cref{eq:fullexpr}.
We substitute the integration variable $t = \tfrac{-i}{\sqrt{2u}}$ and obtain the expression
\begin{equation}
    D_{d, \nu}^{(0)}(\Lambda_{d'}, k, \bm k_\parallel, \bm r)
    = \delta_{l0}\delta_{\bm r 0}
    \int\limits_{-\frac{1}{2\eta^2}}^\infty \dd u\,
    \ee^{-u}
    \begin{cases}
        \frac{\ii}{u} & d = 2\\
        \frac{1}{4u^{\frac{3}{2}}} & d = 3
    \end{cases}
\end{equation}
that can be readily evaluated using the upper incomplete gamma function
\begin{equation}\label{eq:zero}
    D_{d, \nu}^{(0)}(\Lambda_{d'}, k, \bm k_\parallel, \bm r)
    = \delta_{l0}\delta_{\bm r 0}
    \begin{cases}
        \frac{\ii}{\pi} \Gamma\left(0, -\frac{1}{2\eta^2}\right) & d = 2\\
        \frac{1}{4\pi} \Gamma\left(-\frac{1}{2}, -\frac{1}{2\eta^2}\right) & d = 3
    \end{cases}
    \,.
\end{equation}
Note that the conditions on the lower boundary with the particular choice of substitution implies that the upper boundary becomes $\infty$.
Special care has to be taken also for the branch choice in the incomplete gamma function due to its negative argument.
The substitution that was necessary for the integral transformation implies that one has to take the value for $-\tfrac{1}{2\eta^2}-i\epsilon$ for $\epsilon \rightarrow 0^+$, i.e., the value below the real axis.

Having dealt with the origin contribution, we now continue with the main part of the long range summation by the transformation to reciprocal space using the Poisson summation formula
\begin{widetext}
    \begin{align}
        D_{d, \nu}^{(1)}(\Lambda_{d'}, k, \bm k_\parallel, \bm r)
        =
        &
        \frac{2 k^{|l|}}{\ii \pi V_{d'}}
        \sum_{\bm G \in \Lambda_{d'}^\ast}
        \ee^{-\ii (\bm k_\parallel + \bm G) \bm r_\parallel}
        \int\limits_{(0)}^\eta \dd t\,t^{2|l| - 3 + d}
        \ee^{\frac{1}{2t^2}}
        \notag \\
        &\cdot
        \int_{\mathbb{R}^{d'}} \dd^{d'}r' |\bm r' - \bm r_\perp|^{|l|}
        \exp\left(-\frac{(k|\bm r' - \bm r_\perp|t)^2}{2}\right)
        \ee^{-\ii (\bm k_\parallel + \bm G) \bm r'}
        \begin{cases}
            (-1)^\frac{l - |l|}{2} \ee^{\ii l \phi_{\bm r' - \bm r_\perp}} & d = 2\\
            \sqrt{\frac{\pi}{2}} Y_{l m}(\bm r' - \bm r_\perp) & d = 3
        \end{cases}
        \label{eq:recsum}
    \end{align}
\end{widetext}
where we also performed a shift of the newly introduced integral over $\bm r'$ to absorb the component $\bm r_\parallel$ in the integral expression. We observe that components tangential to the lattice enter the expression now only with a phase factor $\ee^{-\ii (\bm k_\parallel + \bm G) \bm r_\parallel}$. Perpendicular shifts with respect to the lattice are considerably more difficult due to the way they appear in \cref{eq:recsum}. The $d'$ dimensional volume of the unit cell is $V_{d'}$.

At this point, it is necessary to individually treat the different cases of $d$ and $d'$.
First, we consider \emph{full} lattices, i.e., lattices where $d = d'$. There, no perpendicular component exists, and the integrals are straightforwardly solved.
However, the cases where $d > d'$ are each solved separately. All possible cases are discussed in following sections.

\subsection{Case: $d = d'$}

The two cases, $d = 2 = d'$, and $d = 3 = d'$, are among the most commonly found ones in literature, and the results are known~\cite{eyert2012}. However, we will rederive them here, since it is instructive to follow the different steps before applying them to the derivation of the more difficult expressions in the other cases.

We focus first on the integration
\begin{equation}
        \int_{\mathbb{R}^d} \dd^d r'\, r'^{|l|}
        \ee^{-\frac{(kr't)^2}{2}}
        \ee^{-\ii (\bm k_\parallel + \bm G) \bm r'}
        \begin{cases}
            \ee^{\ii l \phi_{\bm r'}} & d = 2\\
            Y_{l m}(\bm r') & d = 3
        \end{cases}
\end{equation}
over $\bm r'$. We remark that in those cases, a perpendicular component to the lattice cannot exist, and we set $\bm r_\perp = 0$ in \cref{eq:recsum}.
Using the expansions of the plane wave $\ee^{-\ii (\bm k_\parallel + \bm G) \bm r'}$ suitable for the cases $d = 2$ and $d = 3$ (appendix~\ref{app:pwexp}), we can perform the angular integration trivially due to the orthogonality of the angular functions. The remaining radial integration for the case $d = 2$ is
\begin{equation}
        (-\ii)^{|l|}
        \ee^{\ii l \phi_{\bm k_\parallel + \bm G}}
        \int\limits_0^\infty \dd r' \, r'^{|l| + 1}
        \ee^{-\frac{(kr't)^2}{2}}
        J_{|l|}(\beta k r')
\end{equation}
and 
\begin{equation}
        4 \pi (-\ii)^{l}
        Y_{l m}(\bm k_\parallel + \bm G)
        \int\limits_0^\infty \dd r' \, r'^{|l| + 2}
        \ee^{-\frac{(kr't)^2}{2}}
        j_l (\beta k r')
\end{equation}
for the case $d = 3$, where we introduced $\beta = \tfrac{|\bm k_\parallel + \bm G|}{k}$.
The integral is in both cases essentially the same and can be found in literature~\cite[Eq. 6.631 4.]{gradstejn2014}. Thus, we're now left with
\begin{align}
    \begin{split}
    D_{d, \nu}^{(1)}(\Lambda_d, k, \bm k_\parallel, \bm r)
    =
    \frac{4 (-\ii)^{l}}{\ii V_d k^d}
    \sum_{\bm G \in \Lambda_{d}^\ast}
    \ee^{-\ii (\bm k_\parallel + \bm G) \bm r}
    \beta^{|l|}
    \\
    \cdot
    \int\limits_{(0)}^\eta \frac{\dd t}{t^3}
    \ee^{\frac{\gamma^2}{2t^2}}
    \begin{cases}
        \ee^{\ii l \phi_{\bm k + \bm G}} & d = 2 \\
        \pi Y_{l m}(\bm k_\parallel + \bm G) & d = 3
    \end{cases}
    \,,
    \end{split}
\end{align}
where we use $\gamma = \sqrt{1 - \beta^2}$ with the square root chosen such, that it has a non-negative imaginary part.
The remaining integral over $t$ can be substituted to a simple exponential, that we write here as the incomplete gamma function
\begin{equation}\label{eq:rec:fulllattice}
\begin{split}
    D_{d, \nu}^{(1)}(\Lambda_d, k, \bm k_\parallel, \bm r)
    =
    \frac{4 (-\ii)^{l - 1}}{V_d k^d}
    \sum_{\bm G \in \Lambda_{d}^\ast}
    \ee^{-\ii (\bm k_\parallel + \bm G) \bm r}
    \beta^{|l|}
    \\
    \cdot
    \gamma^{-2} \Gamma\left(1, -\frac{\gamma^2}{2\eta^2}\right)
    \begin{cases}
        \ee^{\ii l \phi_{\bm k + \bm G}} & d = 2 \\
        \pi Y_{l m}(\bm k_\parallel + \bm G) & d = 3
    \end{cases}
    \end{split}
\end{equation}
to highlight the similarities to the following cases.
For the case of $d = d'$, this calculation was quite straightforward compared to the other cases, especially since $\bm r_\perp \neq 0$ is not possible. However, the basic idea of the calculation -- expanding the plane wave suitably and then using a direct evaluation of the integral -- remains the same for $d \neq d'$, although the details become more involved. They will be discussed in the following.

\subsection{Case: $d = 3$, $d' = 2$}
This case has been treated in-depth by Kambe~\cite{kambe1968}, and a direct approach to the solution of this series exists for the case when $\bm r_\perp = 0$~\cite{solbrig1982}.
We now generalize that derivation to the case when $\bm r_\perp \neq 0$. We start with the expression in \cref{eq:recsum}.
Conventionally, we place the lattice in the $z = 0$ plane. By inserting $\bm r_\perp = z \hat{\bm z}$ we obtain
\begin{widetext}
    \begin{equation}
        \label{eq:sw2d:rec:start}
        \begin{split}
        D_{3, l m}^{(1)}(\Lambda_{2}, k, \bm k_\parallel, \bm r)
        =
        & \sqrt{\frac{2}{\pi}}
        \frac{k^{l}}{\ii V_2}
        \sum_{\bm G \in \Lambda_{2}^\ast}
        \ee^{-\ii (\bm k_\parallel + \bm G) \bm r_\parallel}
        \int\limits_{(0)}^\eta \dd t\,t^{2l}
        \ee^{\frac{1}{2t^2}}
        \\
        &\cdot
        \int_{\mathbb{R}^2} \dd^{2}r' (r'^2 + z^2)^{\frac{l}{2}}
        \ee^{-\frac{k^2(r'^2 + z^2)t^2}{2}}
        \ee^{-\ii (\bm k_\parallel + \bm G) \bm r'}
        N_{l |m|} \ee^{\ii m \phi_{\bm r'}}
        (-1)^{\frac{m - |m|}{2}}
        P_l^{|m|} \left(\frac{-z}{\sqrt{r'^2 + z^2}} \right)
        \,,
        \end{split}
    \end{equation}
\end{widetext}
where we have replaced the spherical harmonics with a more explicit expression (\cref{eq:sphharm}).

Now, we replace the plane wave $\ee^{-\ii (\bm k_\parallel + \bm G) \bm r'}$ by a suitable expansion for the evaluation of the spatial integral. The integration domain covers the $d' = 2$ dimensional space and, therefore, the plane wave is expanded in cylindrical coordinates (\cref{eq:pwexp:cw}).
Now, the azimuthal angle integral can be solved trivially, because the phase factors involving $\phi_{\bm r}$ match exactly. The remaining radial integral is
\begin{equation}\label{eq:sw2d:radialint}
    \int\limits_0^\infty \dd r' \, r'
    (r'^2 + z^2)^{\frac{l}{2}}
    J_{|m|}(\beta k r')
    \ee^{-\frac{(k r' t)^2}{2}}
     P_l^{|m|} \left(\frac{-z}{\sqrt{r'^2 + z^2}} \right)
     \,.
\end{equation}
We insert a suitable representation of the Legendre polynomials (\cref{eq:legendrep:closedform}) to eliminate the factor $(r'^2 + z^2)^{\tfrac{l}{2}}$.
Up to a sum over $s \in \{0, 1, \cdots, \lfloor \tfrac{l - |m|}{2} \rfloor\}$ and the prefactors coming from the Legendre polynomial representation.

The integral in \cref{eq:sw2d:radialint} can now be evaluated as
\begin{equation}
    \label{eq:sw2d:int}
    \begin{split}
    \int\limits_0^\infty& \dd r'
    \,r'^{1 + |m| + 2s}
    J_{|m|}(\beta k r')
    \ee^{-\frac{(k r' t)^2}{2}}
    \\
    =&
    \frac{(s + |m|)!}{|m|! \beta k \left(\frac{k^2 t^2}{2}\right)^{\frac{1 + |m|}{2} + s}}
    \ee^{-\frac{\beta^2}{4t^2}}
    M_{\frac{1 + |m|}{2} + s, \frac{|m|}{2}}\left(\frac{\beta^2}{2t^2}\right)
    \\
    =&
    \frac{(s + |m|)!}{\beta k}
    \left(\frac{\beta}{kt^2}\right)^{1 + |m| + 2s}
    (-1)^s \ee^{-\frac{\beta^2}{2t^2}}
    \\
    &\cdot
    \sum_{n = 0}^s
    \binom{s}{n}
    \frac{\left(-\frac{\beta^2}{2t^2}\right)^{-n}}{(s + |m| - n)!}
    \,,
    \end{split}
\end{equation}
where we use the known result of the integral~\cite[Eq. 6.631 1.]{gradstejn2014} and use that $M_{\frac{1 + |m|}{2} + s, \frac{|m|}{2}}(\tfrac{\beta^2}{2t^2})$ is a special case of the Whittaker function that can be expressed as a finite sum of elementary functions (\cref{eq:whittaker}).
Combining \cref{eq:sw2d:rec:start,eq:sw2d:int,eq:legendrep:closedform}, we obtain
\begin{widetext}
    \begin{equation}
        \begin{split}
        D_{3, l m}^{(1)}(\Lambda_2, k, \bm k_\parallel, \bm r)
        =
        \sqrt{2(2l + 1) (l - m)! (l + m)!}
        \frac{\ii^{m - 1}}{V_2 k^2}
        \sum_{\bm G \in \Lambda_{2}^\ast}
        \ee^{-\ii (\bm k_\parallel + \bm G) \bm r_\parallel}
        \ee^{\ii m \phi_{\bm{k}_\parallel + \bm{G}}}
        \int\limits_{(0)}^\eta \dd t\,
        \ee^{\frac{\gamma^2}{2t^2} - \frac{k^2z^2t^2}{2}}
        \\
        \cdot
        \sum_{s = 0}^{\left\lfloor\frac{l - |m|}{2}\right\rfloor}
        \sum_{n = 0}^s
        t^{2l - 2 - 2|m| - 4s + 2n}
        \frac{
            \beta^{|m| + 2s - 2n}
            (-kz)^{l - |m| - 2s}
            (-1)^n
        }{
            2^{2s + |m| - n}
            (s + |m| - n)!n!(s - n)!
            (l - |m| - 2s)!
        }
        \,.
        \end{split}
    \end{equation}
\end{widetext}
The final step is now to simplify the expressions, especially the exponent of $t$, by making it only dependent on the outer sum index to improve the practicality for a software implementation.
Lengthy, but straightforward manipulations of the two nested series (\cref{eq:sumswap:sw2d}) lead to the expression
    \begin{equation}\label{eq:rec:sw2d}
        \begin{split}
        D_{3, l m}^{(1)}(\Lambda_2, k, \bm k_\parallel, \bm r)
        =
        \sum_{\bm{G}\in\Lambda_2^\ast} \ee^{-\ii(\bm{k}_\parallel + \bm{G})\bm{r}}
        \ee^{\ii m \phi_{\bm{k}_\parallel + \bm{G}}}
        \\
        \sum_{n=0}^{l-|m|}
        S_{3, l m n, 2}(k, \beta, z)
        \gamma^{2n-1}
        \int\limits_{-\frac{\gamma^2}{2\eta^2}}^\infty \frac{\dd u}{u}\, u^{\frac{1}{2} - n}
        \ee^{-u+\frac{(\gamma k z)^2}{4u}}
        \end{split}
    \end{equation}
with
    \begin{equation}
        \begin{split}
        S_{3, l m n, 2}(k, \beta, z) 
        =
        \frac{\sqrt{(2l+1)(l-m)!(l+m)!}(-\ii)^{m}}{(-2)^l V_2 k^2}&
        \\
        \sum_{s=n}^{\min(l-|m|,2n)}
        \frac{(-k z)^{2n-s} \beta^{l-s}}
        {(2n-s)!(s-n)!(\frac{l+m-s}{2})!(\frac{l-m-s}{2})!}&
        \,.
        \end{split}
    \end{equation}
For this final expression, we also substitute $t = \tfrac{-\ii \gamma}{\sqrt{2u}}$ which, again, transforms the lower boundary to an integration to infinity.
We emphasize that the sum over $s$ runs only over either all even or all odd values, such that the factorials only take integer values. Thus, $s$ takes on only values with the same parity as $l + m$.
The sum $S_{3, l m n, 2}$ simplifies significantly if $z = 0$, where one gets the simpler expressions from \cref{eq:simplify:s32}.
Now, only the integral for $u$ has to be solved. If $z = 0$, the integral is the upper incomplete gamma function $\Gamma(\tfrac{1}{2} - n, -\tfrac{\gamma^2}{2\eta^2})$, otherwise it can be transformed to an integral $I_l$ (appendix~\ref{app:realrecint}) that we defined already for the real space sum. The appearance of the incomplete gamma function with half integer values is typical for the case $d - d' = 1$ and will later also appear for $d = 2$ and $d' = 1$. When $|\bm r_{\perp}|=0$, our result is equivalent to Kambe's expressions~\cite{kambe1968}.

With our approach working for previously know cases, we now apply it to 1D lattices where a derivation of an equivalent result is not known to us.

\subsection{Case: $d = 3$, $d' = 1$}

Here, we treat the one dimensional lattice in 3D space. We place the lattice along the z-axis of our coordinate systems (\cref{fig:layout}(c)). Then, starting from \cref{eq:recsum}, we can obtain
\begin{widetext}
    \begin{equation}\label{eq:sw1d:start}
        \begin{split}
        D_{3, l m}^{(1)}(\Lambda_1, k, k_\parallel, \bm r)
        =
        &
        \frac{\sqrt{2} k^{l}}{\ii \sqrt{\pi} V_1}
        \sum_{G \in \Lambda_1^\ast}
        \ee^{-\ii (k_\parallel + G) r_\parallel}
        \int\limits_{(0)}^\eta \dd t\,t^{2l}
        \ee^{\frac{1}{2t^2}}
        \\
        &
        \cdot
        \int\limits_{-\infty}^\infty \dd r' (r'^2 + \rho^2)^{\frac{l}{2}}
        \ee^{-\frac{k^2(r'^2 + \rho^2)t^2}{2}}
        \ee^{-\ii (k_\parallel + G) r'}
        N_{l |m|}
        (-1)^{\frac{m - |m|}{2}}
        \ee^{\ii m \phi_{-\bm r_\perp}}
        P_l^{|m|} \left(\frac{r'}{\sqrt{r'^2 + \rho^2}} \right)
        \,,
        \end{split}
    \end{equation}
\end{widetext}
where we used $|\bm r_\perp| = \rho$.
Also, we can now use simple scalars $k_\parallel = \bm k_\parallel \bm{\hat{e}}_z$ and $G = \bm G \bm{\hat{e}}_z$ instead of vectors for the parallel wave vector component and the reciprocal lattice vectors.
For lattices with $d' = 1$, there is no angular integration to do.
We can reuse the expansion of the Legendre polynomials (\cref{eq:legendrep:closedform}) to remove the factor $(r'^2+\rho^2)^\frac{l}{2}$, trading it instead for an additional sum.
After inserting the expansion, we integrate over $r'$ which is, again, an integral that can be found in literature~\cite[3.462 2.]{gradstejn2014}
\begin{align}\label{eq:sw1d:rintegral}
    \begin{split}
        \int\limits_{-\infty}^\infty \dd r'&
        \ee^{-\frac{k^2r'^2t^2}{2}}
        \ee^{-\ii (k_\parallel + G) r'}
        r'^{l - |m| - 2s}
        \\
        =&
        \frac{
            (l - m -2s)!
            \sqrt{2\pi}
        }{
            kt
        }
        \ee^{-\frac{\beta^2}{2t^2}}
        \left(-\frac{\ii\beta}{kt^2}\right)^{l - |m| - 2s}
        \\
        &\cdot
        \sum_{n = 0}^{\left\lfloor\frac{l - |m|}{2} - s\right\rfloor}
        \frac{
            \left(-\frac{t^2}{2\beta^2}\right)^n
        }{
            (l - |m| - 2s -2n)! n!
        }
    \end{split}
\end{align}
and results in a finite series.
Now, we are ready to assemble the full expression
\begin{widetext}
    \begin{equation}
        \begin{split}
        D_{3, l m}^{(1)}(\Lambda_1, k, k_\parallel, \bm r)
        =
        \frac{
            \sqrt{\frac{2l + 1}{\pi}(l - m)!(l + m)!}
        }{
            \ii k V_1
        }
        (-\ii)^{l + m}
        \sum_{G \in \Lambda_1^\ast}
        \ee^{-\ii (k_\parallel + G) r_\parallel}
        \int\limits_{(0)}^\eta \dd t\,
        \ee^{\frac{\gamma^2}{2t^2}-\frac{k^2\rho^2t^2}{2}}
        \ee^{\ii m \phi_{-\bm r_\perp}}
        \\
        \sum_{s = 0}^{\lfloor\frac{l - |m|}{2}\rfloor}
        \sum_{n = 0}^{\left\lfloor\frac{l - |m|}{2} - s\right\rfloor}
        \frac{(k\rho)^{2s + |m|}}{2^{2s + |m| + n}(s + |m|)! s!}
        t^{4s + 2|m| + 2n - 1}
        \beta^{l - |m| - 2s - 2n}
        \frac{
            (-1)^n
        }{
            (l - |m| - 2s -2n)! n!
        }
        \end{split}
    \end{equation}
\end{widetext}
from \cref{eq:sw1d:start,eq:sw1d:rintegral,eq:legendrep:closedform}. Here, we use $\beta = \tfrac{k_\parallel}{k}$, where $k_\parallel$ is the (signed) scalar value of the parallel wave vector component.
We can perform manipulations on the two nested finite series (\cref{eq:sumswap:sw1d}) to finally get the expression
    \begin{equation}\label{eq:rec:sw1d}
        \begin{split}
        D_{3, l m}^{(1)}(\Lambda_1, k,  k_\parallel, \bm r)
        =
        \ee^{\ii m \phi_{-\bm r_\perp}}
        \sum_{G \in \Lambda_1^\ast}
        \ee^{-\ii (k_\parallel + G) r_\parallel}
        \\
        \sum_{n=|m|}^l
        S_{3, l m n, 1}(k, \beta, \rho)
        \frac{\gamma^{2n}}{4^n}
        \int\limits_{-\frac{\gamma^2}{2\eta^2}}^\infty \frac{\dd u}{u}\, u^{-n}
        \ee^{-u+\frac{(\gamma k \rho)^2}{4u}}
        \end{split}
    \end{equation}    
with
    \begin{equation}
        \begin{split}
        S_{3, l m n, 1} (k, \beta, \rho)
        =
        \frac{(-\ii)^{l + 1} \ii^m}{2 V_1 k}
        \sqrt{\frac{2l+1}{\pi}(l-m)!(l+m)!}&
        \\
        \sum_{s=n}^{\min(2n-|m|, l)}
        \frac{(k \rho)^{2n-s} \beta^{l-s}}
        {
            \left(n - \frac{s+m}{2}\right)!
            \left(n - \frac{s-m}{2}\right)!
            (l-s)!
            (s-n)!}&
            \,.
        \end{split}
    \end{equation}
Again, the summation for $s$ only takes values, such that the factorials have an integer argument, namely $s$ must have the same parity as $m$.

As in the previous case, the expression can be simplified significantly (\cref{eq:simplify:s31}) if $\rho = 0$, where the remaining integral transforms to the incomplete gamma function $\Gamma(-n, -\tfrac{\gamma^2}{2\eta^2})$.
If $\rho \neq 0$ the integral can, again, be computed by recurrence (appendix~\ref{app:realrecint}).

\begin{figure*}
\includegraphics[width=\linewidth]{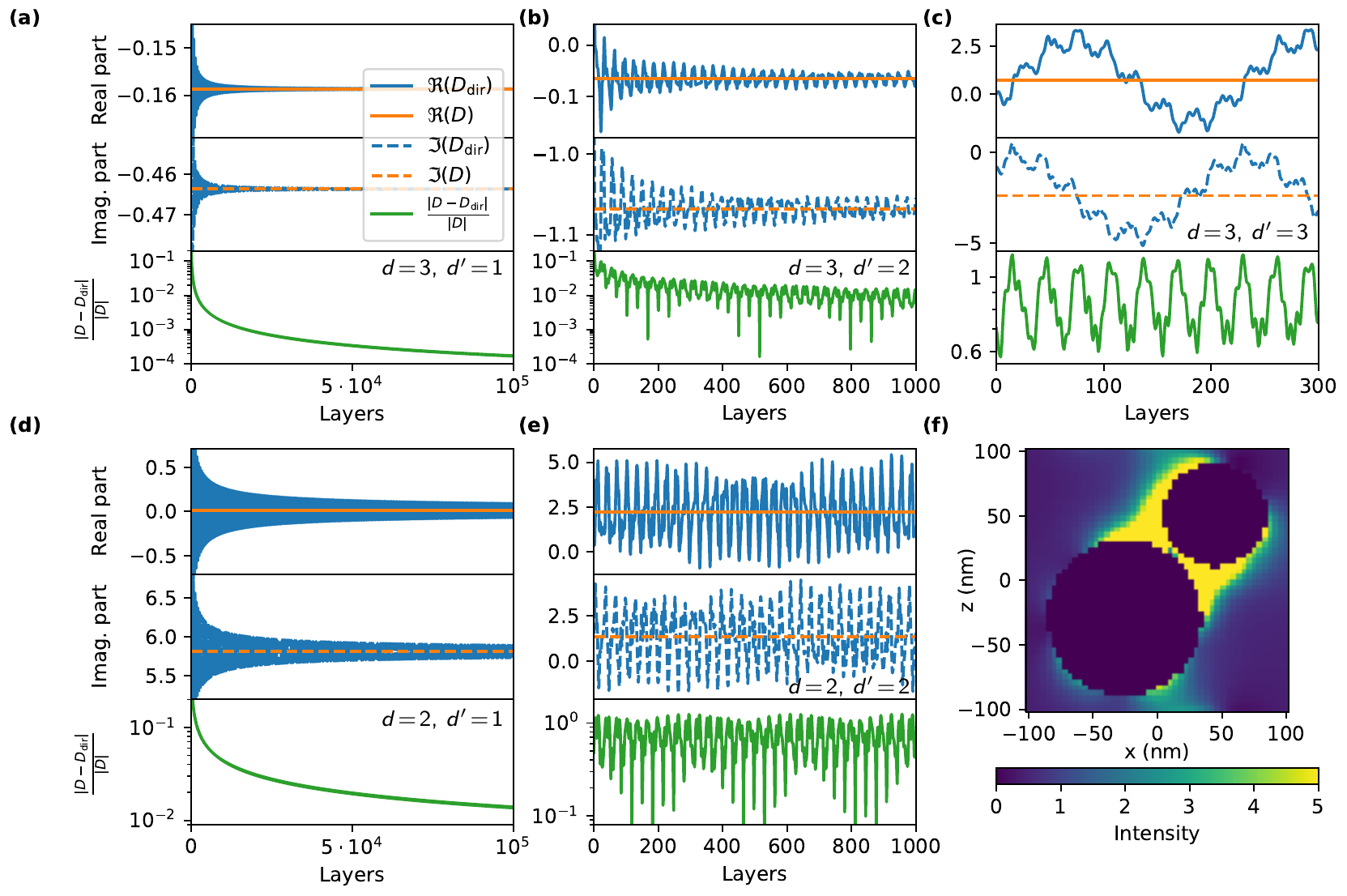}
\caption{Comparison of the direct evaluation of the series and the value for the exponentially convergent expressions. Each panel shows the real part (solid line) and imaginary part (dashed line) of the value for the direct summation (blue), the exponentially convergent ones (orange), and the relative deviation of the direct summation (green). The x-axis shows the number of included layers. These layers have a square and cubic shape for 2D and 3D lattices, respectively. Panels (a), (b), and (c) show the values for the 1D lattice (chain), 2D lattice (grating), and 3D lattice examples for the spherical solution of the Helmholtz equation, respectively. Panels (d) and (e) show the values for the 1D lattice and 2D lattice for the cylindrical solution, respectively. Panel (f) shows an example of an application of the lattice sums for a chain of spheres on two sublattices. The lattice sums are used to calculate the coupling of the spheres within the T-matrix framework and also to compute the electric field whose intensity is shown here.}\label{fig:all}
\end{figure*}

\subsection{Case: $d = 2$, $d' = 1$}
The last case left, treats the lattice sum on a 1D lattice for cylindrical solutions, which we place along the x-axis of our coordinate system (\cref{fig:layout}(e)).
Then, \cref{eq:recsum} becomes
\begin{widetext}
    \begin{equation}
        \begin{split}
        D_{2, l}^{(1)}(\Lambda_1, k, k_\parallel, \bm r)
        =
        \frac{2 k^{|l|} (-1)^\frac{l - |l|}{2}}{\ii \pi V_{1}}
        \sum_{ G \in \Lambda_{1}^\ast}
        \ee^{-\ii (k_\parallel + G) r_\parallel}
        \int\limits_{(0)}^\eta \dd t\,t^{2|l| - 1}
        \ee^{\frac{1}{2t^2}}&
        \\
        \int\limits_{-\infty}^\infty \dd r' (r'^2 + y^2)^{\frac{|l|}{2}}
        \ee^{-\frac{k^2 (r'^2 + y^2) t^2}{2}
        -\ii (k_\parallel + G) r'}
        \left( \frac{r' - \ii \sgn(l) y}{\sqrt{r'^2 + y^2}} \right)^{|l|}&
        \,.
        \end{split}
    \end{equation}
\end{widetext}
Here we used $\bm r_\perp = y\bm{\hat{e}}_y$. The term in brackets to the right corresponds to $\ee^{\ii l \phi_{\bm r' - \bm r_\perp}}$ and its denominator cancels the factor $(r'^2+y^2)^\frac{|l|}{2}$ exactly. We can expand its numerator using the binomial theorem, which replaces it with a sum over $s \in \{0, 1, \cdots, |l|\}$. The spatial integral over $r'$ for each term in the expansion of the binomial is essentially the same as \cref{eq:sw1d:rintegral}
\begin{align}
\begin{split}
    \int\limits_{-\infty}^\infty \dd r'\,
    r'^{s}
    \ee^{-\frac{k r'^2 t^2}{2}}
    \ee^{-\ii (k_\parallel + G) r'}
    \\
    =
    \frac{
        s!
        \sqrt{2\pi}
    }{
        kt
    }
    \ee^{-\frac{\beta^2}{2t^2}}
    \left(-\frac{\ii\beta}{kt^2}\right)^{s}
    \sum_{n = 0}^{\left\lfloor\frac{s}{2}\right\rfloor}
    \frac{
        \left(-\frac{t^2}{2\beta^2}\right)^n
    }{
        (s - 2n)! n!
    }
\end{split}
\end{align}
and can be solved accordingly~\cite[3.462 2.]{gradstejn2014}.
Combining these results, we get
    \begin{equation}
        \begin{split}
        D_{2, l}^{(1)}(\Lambda_1, k, k_\parallel, \bm r)
        =
        \frac{2 \ii^l \sqrt{2}}{\ii \sqrt{\pi} k V_{1}}
        \sum_{ G \in \Lambda_{1}^\ast}
        \ee^{-\ii (k_\parallel + G) r_\parallel}&
        \\
        \int\limits_{(0)}^\eta \dd t\,
        \ee^{\frac{\gamma^2}{2t^2} - \frac{k^2 y^2 t^2}{2}}
        \sum_{s=0}^{|l|}
        \sum_{n = 0}^{\left\lfloor\frac{s}{2}\right\rfloor}
        t^{2(|l| - 1 - s + n)}
        (-1)^{s + n}&
        \\
        \frac{
            |l|!
            (-\sgn(l) k y)^{|l| - s}
            \beta^{s - 2n}
        }{
            (s - 2n)! n! 2^n (|l| - s)!
        }&
        \,.
        \end{split}
    \end{equation}
Similarly to the previous cases, we have two finite series, that can be rearranged to simplify the exponent of the integration variable $t$ (\cref{eq:sumswap:cw1d}), finally arrive at
    \begin{equation}
        \begin{split}
        \label{eq:rec:cw1d}
        D_{2, l}^{(1)}(\Lambda_1, k, k_\parallel, \bm r)
        =
        \sum_{ G \in \Lambda_{1}^\ast}
        \ee^{-\ii (k_\parallel + G) r_\parallel}
        \\
        \sum_{n=0}^{|l|}
        S_{2, l n, 1}(k, \beta, y)
        \gamma^{2n-1}
        \int\limits_{-\frac{\gamma^2}{2\eta^2}}^\infty \frac{\dd u}{u}\, u^{\frac{1}{2} - n}
        \ee^{-u+\frac{(\gamma k y)^2}{4u}}
        \end{split}
    \end{equation}
    with
    \begin{equation}
        \begin{split}
        S_{2, l n, 1} (k, \beta, y)
        =
        \frac{(-\ii)^l 2 }{\sqrt{\pi} V_{1} k}
        &
        \\
        \sum_{s=n}^{\min(2n, |l|)}
        \frac{(-\sgn(l) k y)^{2n-s} \beta^{|l|-s}}
        {
            2^s
            (2n - s)!
            (|l|-s)!
            (s-n)!
        }&
        \end{split}
    \end{equation}
again after substituting $t = \tfrac{-\ii \gamma}{\sqrt{2u}}$.
Here, the sum in $s$ takes every value in its range in contrast to the other cases. Major simplifications are possible when considering $\rho = 0$ (\cref{eq:simplify:s21}), where the integral becomes, analogously to the $d = 3$, $d' = 2$ case, the incomplete gamma function $\Gamma(\tfrac{1}{2} - n, -\tfrac{\gamma^2}{2\eta^2})$.

\section{Comparison with the direct sum} \label{sec:comparison}

We verify and compare our results by evaluating the sum directly with an increasing number of lattice points and by using the novel expressions derived in this work.

In the first example, we use the values $l = 2$ and $m = 0$ in case of $d = 3$ and for $m = 2$ in case of $d = 2$. The shift vector is $\bm r = (0.2, 0.1, 0.3)$ for $d = 3$ and $\bm r = (0.1, 0.3)$ for $d = 2$. The parallel component of the wave vector is $k_\parallel = 0.3$, $\bm k_\parallel = (-0.1, 0.2)$, and $\bm k_\parallel = (0.3, -0.1, 0.2)$ for the 1D, 2D, and 3D lattices, respectively. In all cases, we use $k = 3$ and a lattice pitch $a = 1.9$. The 2D and 3D lattice are square or, suitably, cubic.

For the chain there is mostly only one way to include lattice points in the direct summation, namely taking the origin unit cell and then expanding outwards on both sides. This summation scheme can be generalized to higher dimensions in a spherical or cubic fashion. This means that all points within a region defined by either a fixed Euclidean distance or a fixed Chebyshev distance from the reference unit cell are considered in the sum. We opt for the latter because of its better convergence behavior~\cite{linton2010} and express the number of points considered by the number of layers $n$, i.e., all points with $\lVert \bm R \rVert_\infty \leq n a$.

\begin{figure*}
\includegraphics[width=\linewidth]{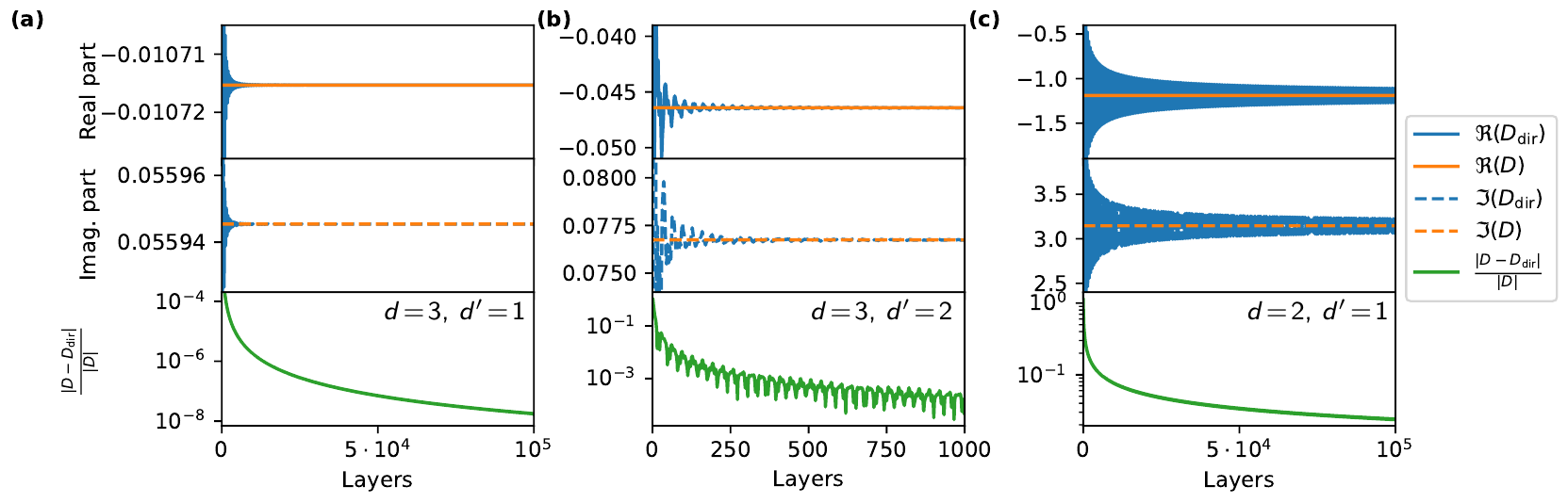}
\caption{Comparison of the direct summation and the value of exponentially fast converging series for large shifts $\bm r_\perp$ perpendicular to the lattice. Panel (a) and (b) shown the results for a 1D and 2D lattice summing spherical solutions. The values of $l = 2$ and $m = 1$ are chosen, which in case of $\bm r_\perp = 0$ must vanish. Panel (c) shows the case $d = 2$ and $d' = 1$.}\label{fig:largeperp}
\end{figure*}

The first row in \cref{fig:all} shows the results for $d = 3$ and $d' \in \{1, 2, 3\}$ in panels (a), (b), and (c), respectively. The second row shows the results for $d = 2$ and $d' \in \{1, 2\}$.
We observe in all five cases presented in \cref{fig:all}(a)-(e) very fast oscillations of the direct sum, depending on the number of layers. For panels (a) and (d), corresponding to the chain, we included up to $10^5$ layers, such that the fast oscillations are not resolved and appear as a blue area. Although converging fairly rapid initially, the direct summation needs more than $10^5$ layers to deviate only by $10^{-4}$ from the Ewald's method result for $d = 3$ and $d' = 1$. For $d = 2$ and $d' = 1$ the convergence is even slower by up to two orders of magnitude.
The exponentially fast converging result obtained with \cref{eq:realsum:sw,eq:realsum:cw} for the real part and \cref{eq:rec:fulllattice,eq:rec:sw2d,eq:rec:sw1d,eq:rec:cw1d} for the reciprocal part is shown as a orange line.

Panel (b) shows the results for $d = 3$ and $d' = 2$. Here, we included up to 1000 layers. In comparison to $d' = 1$, the convergence of the direct sum is more time consuming. To reach a relative accuracy of roughly $10^{-2}$, the contributions of over 4 million lattice points have to be evaluated. Such a deteriorating convergence behavior as $d'$ gets closer to $d$ is commonly found for direct summations. This is even more pronounced in panels (c) and (e), which show the results for the full lattices. Here, for the number of layers included, no convergence is visible at all but only oscillations

In summary, we find that the formulas derived converge quickly to a precise value suitable for numerical evaluations. While it is possible to improve the convergence of the direct summation by averaging over one or multiple oscillations (see appendix~\ref{app:convolution}), it is clear that the exponentially fast converging series are a major improvement.

Additionally to our first example, we also look into larger shifts $|\bm r_\perp|$ away from the lattice and mainly the components that are not present for $\bm r_\perp = 0$. For this, we choose the parameters $l = 2$ and $m = 1$ for $d = 3$ and $m = 2$ for $d = 2$. For $d = 3$ and $d' = 2$ we set $\bm r = (1.5, 1.1, 0.3)$, for $d = 3$ and $d' = 1$ we set $\bm r = (0.2, 0.1, 1.3)$, and for $d = 2$ and $d' = 1$ we set $\bm r = (0.1, 1.3)$. The values of $k$, $a$, and $\bm k_\parallel$ remain unchanged.
The entries with $l = 2$ and $m = 1$ for $d = 3$ that are shown in panels (a) and (b) in \cref{fig:largeperp} would be zero in case of $\bm r_\perp = 0$, but it becomes non-zero when $\bm r_\perp \neq 0$. The direct summation in those cases converges comparably fast, and we can confirm that the derived formulas are correct. Panel (c) shows the result for $d = 2$ and $d' = 1$ with large $\bm r_\perp$. Also, in that case, the result of the derived exponentially convergent formula is approached by the direct summation with an increasing number of layers. However, the convergence is quite slow.

\section{Exemplary application}\label{sec:example}

A typical field of application for the lattice sum is in summing translation coefficients for vector spherical waves like they appear as part of the T-matrix method. Here, we apply the summation for a 1D lattice for vector spherical waves. The example system corresponds roughly to the sketch shown in \cref{fig:layout}(a). It consists of two spheres per unit cell with radii \SI{40}{\nano\metre} and \SI{60}{\nano\metre} and relative permittivity $\epsilon = 9$ with a relative shift $\bm{r} = (70, 0, 80)^T\,\si{\nano\metre}$ from the larger to the smaller sphere. The chain has a lattice constant of \SI{200}{\nano\metre} and is illuminated with a plane wave of wavelength \SI{500}{\nano\metre} under oblique incidence with a $\tfrac{\pi}{6}$ angle with respect to the x-axis.

\Cref{fig:all}(f) shows the field intensity in one unit cell. Using the T-matrix method together with the lattice sums, we can efficiently compute the electric field in the entire space outside the spheres. This example makes not only use of the lattice sums to translate the scattered fields between the two sublattices associated with each type of sphere for computing the mutual interaction but also uses them to translate the scattered field to each point in the sampled space to obtain the electric field within the unit cell.

\section{Conclusion}
We presented a derivation of exponentially fast converging series for quasi-periodic Helmholtz equation sums in $d = 2$ and $d = 3$ spatial dimensions. Our approach is suitable to derive exponentially convergent series for arbitrary lattice dimensions $d' \leq d$. A special emphasis is placed on the applicability of the lattice sums to the case when there is a relative shift between multiple sublattices. This enables us to apply the sums to a wide range of applications.

For an implementation of the exponentially fast lattice sum, the formulas in \cref{eq:realsum:cw,eq:realsum:sw,eq:rec:fulllattice,eq:rec:sw2d,eq:rec:sw1d,eq:rec:cw1d,eq:zero}
can be directly used with the integrals evaluated by recursion.

\begin{acknowledgments}
D.B. and C.R. acknowledge support by the Deutsche Forschungsgemeinschaft (DFG, German Research Foundation) under Germany’s Excellence Strategy via the Excellence Cluster 3D Matter Made to Order (EXC--2082/1--390761711) and from the Carl Zeiss Foundation via CZF-Focus@HEiKA.
\end{acknowledgments}

\bibliography{lattice_sums}

\begin{thebibliography}{48}%
\makeatletter
\providecommand \@ifxundefined [1]{%
 \@ifx{#1\undefined}
}%
\providecommand \@ifnum [1]{%
 \ifnum #1\expandafter \@firstoftwo
 \else \expandafter \@secondoftwo
 \fi
}%
\providecommand \@ifx [1]{%
 \ifx #1\expandafter \@firstoftwo
 \else \expandafter \@secondoftwo
 \fi
}%
\providecommand \natexlab [1]{#1}%
\providecommand \enquote  [1]{``#1''}%
\providecommand \bibnamefont  [1]{#1}%
\providecommand \bibfnamefont [1]{#1}%
\providecommand \citenamefont [1]{#1}%
\providecommand \href@noop [0]{\@secondoftwo}%
\providecommand \href [0]{\begingroup \@sanitize@url \@href}%
\providecommand \@href[1]{\@@startlink{#1}\@@href}%
\providecommand \@@href[1]{\endgroup#1\@@endlink}%
\providecommand \@sanitize@url [0]{\catcode `\\12\catcode `\$12\catcode
  `\&12\catcode `\#12\catcode `\^12\catcode `\_12\catcode `\%12\relax}%
\providecommand \@@startlink[1]{}%
\providecommand \@@endlink[0]{}%
\providecommand \url  [0]{\begingroup\@sanitize@url \@url }%
\providecommand \@url [1]{\endgroup\@href {#1}{\urlprefix }}%
\providecommand \urlprefix  [0]{URL }%
\providecommand \Eprint [0]{\href }%
\providecommand \doibase [0]{https://doi.org/}%
\providecommand \selectlanguage [0]{\@gobble}%
\providecommand \bibinfo  [0]{\@secondoftwo}%
\providecommand \bibfield  [0]{\@secondoftwo}%
\providecommand \translation [1]{[#1]}%
\providecommand \BibitemOpen [0]{}%
\providecommand \bibitemStop [0]{}%
\providecommand \bibitemNoStop [0]{.\EOS\space}%
\providecommand \EOS [0]{\spacefactor3000\relax}%
\providecommand \BibitemShut  [1]{\csname bibitem#1\endcsname}%
\let\auto@bib@innerbib\@empty
\bibitem [{\citenamefont {Varadan}\ and\ \citenamefont
  {Varadan}(1980)}]{varadan1980}%
  \BibitemOpen
  \bibfield  {author} {\bibinfo {author} {\bibfnamefont {V.~K.}\ \bibnamefont
  {Varadan}}\ and\ \bibinfo {author} {\bibfnamefont {V.~V.}\ \bibnamefont
  {Varadan}},\ }\bibfield  {title} {\bibinfo {title} {Acoustic, electromagnetic
  and elastic wave scattering--focus on the {{T-matrix}} approach}\ }(\bibinfo
  {publisher} {{Pergamon Press}},\ \bibinfo {address} {{New York}},\ \bibinfo
  {year} {1980})\BibitemShut {NoStop}%
\bibitem [{\citenamefont {Waterman}(2009)}]{waterman2009}%
  \BibitemOpen
  \bibfield  {author} {\bibinfo {author} {\bibfnamefont {P.~C.}\ \bibnamefont
  {Waterman}},\ }\bibfield  {title} {\bibinfo {title} {T-matrix methods in
  acoustic scattering},\ }\href {https://doi.org/10.1121/1.3035839} {\bibfield
  {journal} {\bibinfo  {journal} {The Journal of the Acoustical Society of
  America}\ }\textbf {\bibinfo {volume} {125}},\ \bibinfo {pages} {42}
  (\bibinfo {year} {2009})}\BibitemShut {NoStop}%
\bibitem [{\citenamefont {Ewald}(1921)}]{ewald1921}%
  \BibitemOpen
  \bibfield  {author} {\bibinfo {author} {\bibfnamefont {P.~P.}\ \bibnamefont
  {Ewald}},\ }\bibfield  {title} {\bibinfo {title} {Die {{Berechnung}}
  optischer und elektrostatischer {{Gitterpotentiale}}},\ }\href
  {https://doi.org/10.1002/andp.19213690304} {\bibfield  {journal} {\bibinfo
  {journal} {Annalen der Physik}\ }\textbf {\bibinfo {volume} {369}},\ \bibinfo
  {pages} {253} (\bibinfo {year} {1921})}\BibitemShut {NoStop}%
\bibitem [{\citenamefont {Babicheva}\ and\ \citenamefont
  {Evlyukhin}(2021)}]{babicheva2021}%
  \BibitemOpen
  \bibfield  {author} {\bibinfo {author} {\bibfnamefont {V.~E.}\ \bibnamefont
  {Babicheva}}\ and\ \bibinfo {author} {\bibfnamefont {A.~B.}\ \bibnamefont
  {Evlyukhin}},\ }\bibfield  {title} {\bibinfo {title} {Multipole lattice
  effects in high refractive index metasurfaces},\ }\href
  {https://doi.org/10.1063/5.0024274} {\bibfield  {journal} {\bibinfo
  {journal} {Journal of Applied Physics}\ }\textbf {\bibinfo {volume} {129}},\
  \bibinfo {pages} {040902} (\bibinfo {year} {2021})}\BibitemShut {NoStop}%
\bibitem [{\citenamefont {Berkhout}\ and\ \citenamefont
  {Koenderink}(2020)}]{berkhout2020}%
  \BibitemOpen
  \bibfield  {author} {\bibinfo {author} {\bibfnamefont {A.}~\bibnamefont
  {Berkhout}}\ and\ \bibinfo {author} {\bibfnamefont {A.~F.}\ \bibnamefont
  {Koenderink}},\ }\bibfield  {title} {\bibinfo {title} {A simple
  transfer-matrix model for metasurface multilayer systems},\ }\href
  {https://doi.org/10.1515/nanoph-2020-0212} {\bibfield  {journal} {\bibinfo
  {journal} {Nanophotonics}\ }\textbf {\bibinfo {volume} {9}},\ \bibinfo
  {pages} {3985} (\bibinfo {year} {2020})}\BibitemShut {NoStop}%
\bibitem [{\citenamefont {Chen}\ \emph {et~al.}(2017)\citenamefont {Chen},
  \citenamefont {Zhang},\ and\ \citenamefont {Koenderink}}]{chen2017}%
  \BibitemOpen
  \bibfield  {author} {\bibinfo {author} {\bibfnamefont {Y.}~\bibnamefont
  {Chen}}, \bibinfo {author} {\bibfnamefont {Y.}~\bibnamefont {Zhang}},\ and\
  \bibinfo {author} {\bibfnamefont {A.~F.}\ \bibnamefont {Koenderink}},\
  }\bibfield  {title} {\bibinfo {title} {General point dipole theory for
  periodic metasurfaces: Magnetoelectric scattering lattices coupled to planar
  photonic structures},\ }\href {https://doi.org/10.1364/OE.25.021358}
  {\bibfield  {journal} {\bibinfo  {journal} {Optics Express}\ }\textbf
  {\bibinfo {volume} {25}},\ \bibinfo {pages} {21358} (\bibinfo {year}
  {2017})}\BibitemShut {NoStop}%
\bibitem [{\citenamefont {Cummins}\ \emph {et~al.}(1976)\citenamefont
  {Cummins}, \citenamefont {Dunmur}, \citenamefont {Munn},\ and\ \citenamefont
  {Newham}}]{cummins1976}%
  \BibitemOpen
  \bibfield  {author} {\bibinfo {author} {\bibfnamefont {P.~G.}\ \bibnamefont
  {Cummins}}, \bibinfo {author} {\bibfnamefont {D.~A.}\ \bibnamefont {Dunmur}},
  \bibinfo {author} {\bibfnamefont {R.~W.}\ \bibnamefont {Munn}},\ and\
  \bibinfo {author} {\bibfnamefont {R.~J.}\ \bibnamefont {Newham}},\ }\bibfield
   {title} {\bibinfo {title} {Applications of the {{Ewald}} method. {{I}}.
  {{Calculation}} of multipole lattice sums},\ }\href
  {https://doi.org/10.1107/S0567739476001708} {\bibfield  {journal} {\bibinfo
  {journal} {Acta Crystallographica Section A: Crystal Physics, Diffraction,
  Theoretical and General Crystallography}\ }\textbf {\bibinfo {volume} {32}},\
  \bibinfo {pages} {847} (\bibinfo {year} {1976})}\BibitemShut {NoStop}%
\bibitem [{\citenamefont {Gallinet}\ \emph {et~al.}(2010)\citenamefont
  {Gallinet}, \citenamefont {Kern},\ and\ \citenamefont
  {Martin}}]{gallinet2010}%
  \BibitemOpen
  \bibfield  {author} {\bibinfo {author} {\bibfnamefont {B.}~\bibnamefont
  {Gallinet}}, \bibinfo {author} {\bibfnamefont {A.~M.}\ \bibnamefont {Kern}},\
  and\ \bibinfo {author} {\bibfnamefont {O.~J.~F.}\ \bibnamefont {Martin}},\
  }\bibfield  {title} {\bibinfo {title} {Accurate and versatile modeling of
  electromagnetic scattering on periodic nanostructures with a surface integral
  approach},\ }\href {https://doi.org/10.1364/JOSAA.27.002261} {\bibfield
  {journal} {\bibinfo  {journal} {JOSA A}\ }\textbf {\bibinfo {volume} {27}},\
  \bibinfo {pages} {2261} (\bibinfo {year} {2010})}\BibitemShut {NoStop}%
\bibitem [{\citenamefont {Goodarzi}\ and\ \citenamefont
  {Pakizeh}(2021)}]{goodarzi2021}%
  \BibitemOpen
  \bibfield  {author} {\bibinfo {author} {\bibfnamefont {M.}~\bibnamefont
  {Goodarzi}}\ and\ \bibinfo {author} {\bibfnamefont {T.}~\bibnamefont
  {Pakizeh}},\ }\bibfield  {title} {\bibinfo {title} {Retrieving effective
  surface susceptibilities of high-index metasurfaces based on dipole
  approximation},\ }\href {https://doi.org/10.1016/j.optcom.2020.126659}
  {\bibfield  {journal} {\bibinfo  {journal} {Optics Communications}\ }\textbf
  {\bibinfo {volume} {483}},\ \bibinfo {pages} {126659} (\bibinfo {year}
  {2021})}\BibitemShut {NoStop}%
\bibitem [{\citenamefont {Hu}\ \emph {et~al.}(2021)\citenamefont {Hu},
  \citenamefont {Wang}, \citenamefont {Mazor}, \citenamefont {Qiu},\ and\
  \citenamefont {Al{\`u}}}]{hu2021}%
  \BibitemOpen
  \bibfield  {author} {\bibinfo {author} {\bibfnamefont {G.}~\bibnamefont
  {Hu}}, \bibinfo {author} {\bibfnamefont {M.}~\bibnamefont {Wang}}, \bibinfo
  {author} {\bibfnamefont {Y.}~\bibnamefont {Mazor}}, \bibinfo {author}
  {\bibfnamefont {C.-W.}\ \bibnamefont {Qiu}},\ and\ \bibinfo {author}
  {\bibfnamefont {A.}~\bibnamefont {Al{\`u}}},\ }\bibfield  {title} {\bibinfo
  {title} {Tailoring {{Light}} with {{Layered}} and {{Moir\'e Metasurfaces}}},\
  }\href {https://doi.org/10.1016/j.trechm.2021.02.004} {\bibfield  {journal}
  {\bibinfo  {journal} {Trends in Chemistry}\ }\textbf {\bibinfo {volume}
  {3}},\ \bibinfo {pages} {342} (\bibinfo {year} {2021})}\BibitemShut {NoStop}%
\bibitem [{\citenamefont {Lovat}\ \emph {et~al.}(2008)\citenamefont {Lovat},
  \citenamefont {Burghignoli},\ and\ \citenamefont {Araneo}}]{lovat2008}%
  \BibitemOpen
  \bibfield  {author} {\bibinfo {author} {\bibfnamefont {G.}~\bibnamefont
  {Lovat}}, \bibinfo {author} {\bibfnamefont {P.}~\bibnamefont {Burghignoli}},\
  and\ \bibinfo {author} {\bibfnamefont {R.}~\bibnamefont {Araneo}},\
  }\bibfield  {title} {\bibinfo {title} {Efficient {{Evaluation}} of the 3-{{D
  Periodic Green}}'s {{Function Through}} the {{Ewald Method}}},\ }\href
  {https://doi.org/10.1109/TMTT.2008.2002232} {\bibfield  {journal} {\bibinfo
  {journal} {IEEE Transactions on Microwave Theory and Techniques}\ }\textbf
  {\bibinfo {volume} {56}},\ \bibinfo {pages} {2069} (\bibinfo {year}
  {2008})}\BibitemShut {NoStop}%
\bibitem [{\citenamefont {Lunnemann}\ \emph {et~al.}(2013)\citenamefont
  {Lunnemann}, \citenamefont {Sersic},\ and\ \citenamefont
  {Koenderink}}]{lunnemann2013}%
  \BibitemOpen
  \bibfield  {author} {\bibinfo {author} {\bibfnamefont {P.}~\bibnamefont
  {Lunnemann}}, \bibinfo {author} {\bibfnamefont {I.}~\bibnamefont {Sersic}},\
  and\ \bibinfo {author} {\bibfnamefont {A.~F.}\ \bibnamefont {Koenderink}},\
  }\bibfield  {title} {\bibinfo {title} {Optical properties of two-dimensional
  magnetoelectric point scattering lattices},\ }\href
  {https://doi.org/10.1103/PhysRevB.88.245109} {\bibfield  {journal} {\bibinfo
  {journal} {Physical Review B}\ }\textbf {\bibinfo {volume} {88}},\ \bibinfo
  {pages} {245109} (\bibinfo {year} {2013})}\BibitemShut {NoStop}%
\bibitem [{\citenamefont {Lunnemann}\ and\ \citenamefont
  {Koenderink}(2016)}]{lunnemann2016}%
  \BibitemOpen
  \bibfield  {author} {\bibinfo {author} {\bibfnamefont {P.}~\bibnamefont
  {Lunnemann}}\ and\ \bibinfo {author} {\bibfnamefont {A.~F.}\ \bibnamefont
  {Koenderink}},\ }\bibfield  {title} {\bibinfo {title} {The local density of
  optical states of a metasurface},\ }\href {https://doi.org/10.1038/srep20655}
  {\bibfield  {journal} {\bibinfo  {journal} {Scientific Reports}\ }\textbf
  {\bibinfo {volume} {6}},\ \bibinfo {pages} {20655} (\bibinfo {year}
  {2016})}\BibitemShut {NoStop}%
\bibitem [{\citenamefont {Rahimzadegan}\ \emph {et~al.}(2022)\citenamefont
  {Rahimzadegan}, \citenamefont {Karamanos}, \citenamefont {Alaee},
  \citenamefont {Lamprianidis}, \citenamefont {Beutel}, \citenamefont {Boyd},\
  and\ \citenamefont {Rockstuhl}}]{rahimzadegan2022}%
  \BibitemOpen
  \bibfield  {author} {\bibinfo {author} {\bibfnamefont {A.}~\bibnamefont
  {Rahimzadegan}}, \bibinfo {author} {\bibfnamefont {T.~D.}\ \bibnamefont
  {Karamanos}}, \bibinfo {author} {\bibfnamefont {R.}~\bibnamefont {Alaee}},
  \bibinfo {author} {\bibfnamefont {A.~G.}\ \bibnamefont {Lamprianidis}},
  \bibinfo {author} {\bibfnamefont {D.}~\bibnamefont {Beutel}}, \bibinfo
  {author} {\bibfnamefont {R.~W.}\ \bibnamefont {Boyd}},\ and\ \bibinfo
  {author} {\bibfnamefont {C.}~\bibnamefont {Rockstuhl}},\ }\bibfield  {title}
  {\bibinfo {title} {A {{Comprehensive Multipolar Theory}} for {{Periodic
  Metasurfaces}}},\ }\href {https://doi.org/10.1002/adom.202102059} {\bibfield
  {journal} {\bibinfo  {journal} {Advanced Optical Materials}\ }\textbf
  {\bibinfo {volume} {10}},\ \bibinfo {pages} {2102059} (\bibinfo {year}
  {2022})}\BibitemShut {NoStop}%
\bibitem [{\citenamefont {Rider}\ \emph {et~al.}(2022)\citenamefont {Rider},
  \citenamefont {Buend{\'i}a}, \citenamefont {Abujetas}, \citenamefont
  {Huidobro}, \citenamefont {{S{\'a}nchez-Gil}},\ and\ \citenamefont
  {Giannini}}]{rider2022}%
  \BibitemOpen
  \bibfield  {author} {\bibinfo {author} {\bibfnamefont {M.~S.}\ \bibnamefont
  {Rider}}, \bibinfo {author} {\bibfnamefont {{\'A}.}~\bibnamefont
  {Buend{\'i}a}}, \bibinfo {author} {\bibfnamefont {D.~R.}\ \bibnamefont
  {Abujetas}}, \bibinfo {author} {\bibfnamefont {P.~A.}\ \bibnamefont
  {Huidobro}}, \bibinfo {author} {\bibfnamefont {J.~A.}\ \bibnamefont
  {{S{\'a}nchez-Gil}}},\ and\ \bibinfo {author} {\bibfnamefont
  {V.}~\bibnamefont {Giannini}},\ }\bibfield  {title} {\bibinfo {title}
  {Advances and {{Prospects}} in {{Topological Nanoparticle Photonics}}},\
  }\href {https://doi.org/10.1021/acsphotonics.1c01874} {\bibfield  {journal}
  {\bibinfo  {journal} {ACS Photonics}\ }\textbf {\bibinfo {volume} {9}},\
  \bibinfo {pages} {1483} (\bibinfo {year} {2022})}\BibitemShut {NoStop}%
\bibitem [{\citenamefont {Stefanou}\ \emph {et~al.}(1998)\citenamefont
  {Stefanou}, \citenamefont {Yannopapas},\ and\ \citenamefont
  {Modinos}}]{stefanou1998}%
  \BibitemOpen
  \bibfield  {author} {\bibinfo {author} {\bibfnamefont {N.}~\bibnamefont
  {Stefanou}}, \bibinfo {author} {\bibfnamefont {V.}~\bibnamefont
  {Yannopapas}},\ and\ \bibinfo {author} {\bibfnamefont {A.}~\bibnamefont
  {Modinos}},\ }\bibfield  {title} {\bibinfo {title} {Heterostructures of
  photonic crystals: Frequency bands and transmission coefficients},\ }\href
  {https://doi.org/10.1016/s0010-4655(98)00060-5} {\bibfield  {journal}
  {\bibinfo  {journal} {Computer Physics Communications}\ }\textbf {\bibinfo
  {volume} {113}},\ \bibinfo {pages} {49} (\bibinfo {year} {1998})}\BibitemShut
  {NoStop}%
\bibitem [{\citenamefont {Stefanou}\ \emph {et~al.}(2000)\citenamefont
  {Stefanou}, \citenamefont {Yannopapas},\ and\ \citenamefont
  {Modinos}}]{stefanou2000}%
  \BibitemOpen
  \bibfield  {author} {\bibinfo {author} {\bibfnamefont {N.}~\bibnamefont
  {Stefanou}}, \bibinfo {author} {\bibfnamefont {V.}~\bibnamefont
  {Yannopapas}},\ and\ \bibinfo {author} {\bibfnamefont {A.}~\bibnamefont
  {Modinos}},\ }\bibfield  {title} {\bibinfo {title} {{{MULTEM}} 2: {{A}} new
  version of the program for transmission and band-structure calculations of
  photonic crystals},\ }\href {https://doi.org/10.1016/s0010-4655(00)00131-4}
  {\bibfield  {journal} {\bibinfo  {journal} {Computer Physics Communications}\
  }\textbf {\bibinfo {volume} {132}},\ \bibinfo {pages} {189} (\bibinfo {year}
  {2000})}\BibitemShut {NoStop}%
\bibitem [{\citenamefont {Yermakov}\ \emph {et~al.}(2018)\citenamefont
  {Yermakov}, \citenamefont {Permyakov}, \citenamefont {Porubaev},
  \citenamefont {Dmitriev}, \citenamefont {Samusev}, \citenamefont {Iorsh},
  \citenamefont {Malureanu}, \citenamefont {Lavrinenko},\ and\ \citenamefont
  {Bogdanov}}]{yermakov2018}%
  \BibitemOpen
  \bibfield  {author} {\bibinfo {author} {\bibfnamefont {O.~Y.}\ \bibnamefont
  {Yermakov}}, \bibinfo {author} {\bibfnamefont {D.~V.}\ \bibnamefont
  {Permyakov}}, \bibinfo {author} {\bibfnamefont {F.~V.}\ \bibnamefont
  {Porubaev}}, \bibinfo {author} {\bibfnamefont {P.~A.}\ \bibnamefont
  {Dmitriev}}, \bibinfo {author} {\bibfnamefont {A.~K.}\ \bibnamefont
  {Samusev}}, \bibinfo {author} {\bibfnamefont {I.~V.}\ \bibnamefont {Iorsh}},
  \bibinfo {author} {\bibfnamefont {R.}~\bibnamefont {Malureanu}}, \bibinfo
  {author} {\bibfnamefont {A.~V.}\ \bibnamefont {Lavrinenko}},\ and\ \bibinfo
  {author} {\bibfnamefont {A.~A.}\ \bibnamefont {Bogdanov}},\ }\bibfield
  {title} {\bibinfo {title} {Effective surface conductivity of optical
  hyperbolic metasurfaces: From far-field characterization to surface wave
  analysis},\ }\href {https://doi.org/10.1038/s41598-018-32479-y} {\bibfield
  {journal} {\bibinfo  {journal} {Scientific Reports}\ }\textbf {\bibinfo
  {volume} {8}},\ \bibinfo {pages} {14135} (\bibinfo {year}
  {2018})}\BibitemShut {NoStop}%
\bibitem [{\citenamefont {Moroz}(2006)}]{moroz2006}%
  \BibitemOpen
  \bibfield  {author} {\bibinfo {author} {\bibfnamefont {A.}~\bibnamefont
  {Moroz}},\ }\bibfield  {title} {\bibinfo {title} {Quasi-periodic {{Green}}'s
  functions of the {{Helmholtz}} and {{Laplace}} equations},\ }\href
  {https://doi.org/10.1088/0305-4470/39/36/009} {\bibfield  {journal} {\bibinfo
   {journal} {Journal of Physics A: Mathematical and General}\ }\textbf
  {\bibinfo {volume} {39}},\ \bibinfo {pages} {11247} (\bibinfo {year}
  {2006})}\BibitemShut {NoStop}%
\bibitem [{\citenamefont {Linton}(2010)}]{linton2010}%
  \BibitemOpen
  \bibfield  {author} {\bibinfo {author} {\bibfnamefont {C.~M.}\ \bibnamefont
  {Linton}},\ }\bibfield  {title} {\bibinfo {title} {Lattice {{Sums}} for the
  {{Helmholtz Equation}}},\ }\href {https://doi.org/10.1137/09075130X}
  {\bibfield  {journal} {\bibinfo  {journal} {SIAM Review}\ }\textbf {\bibinfo
  {volume} {52}},\ \bibinfo {pages} {630} (\bibinfo {year} {2010})}\BibitemShut
  {NoStop}%
\bibitem [{\citenamefont {Capolino}\ \emph {et~al.}(2005)\citenamefont
  {Capolino}, \citenamefont {Wilton},\ and\ \citenamefont
  {Johnson}}]{capolino2005}%
  \BibitemOpen
  \bibfield  {author} {\bibinfo {author} {\bibfnamefont {F.}~\bibnamefont
  {Capolino}}, \bibinfo {author} {\bibfnamefont {D.}~\bibnamefont {Wilton}},\
  and\ \bibinfo {author} {\bibfnamefont {W.}~\bibnamefont {Johnson}},\
  }\bibfield  {title} {\bibinfo {title} {Efficient computation of the 2-{{D
  Green}}'s function for 1-{{D}} periodic structures using the {{Ewald}}
  method},\ }\href {https://doi.org/10.1109/TAP.2005.854556} {\bibfield
  {journal} {\bibinfo  {journal} {IEEE Transactions on Antennas and
  Propagation}\ }\textbf {\bibinfo {volume} {53}},\ \bibinfo {pages} {2977}
  (\bibinfo {year} {2005})}\BibitemShut {NoStop}%
\bibitem [{\citenamefont {Capolino}\ \emph {et~al.}(2007)\citenamefont
  {Capolino}, \citenamefont {Wilton},\ and\ \citenamefont
  {Johnson}}]{capolino2007}%
  \BibitemOpen
  \bibfield  {author} {\bibinfo {author} {\bibfnamefont {F.}~\bibnamefont
  {Capolino}}, \bibinfo {author} {\bibfnamefont {D.~R.}\ \bibnamefont
  {Wilton}},\ and\ \bibinfo {author} {\bibfnamefont {W.~A.}\ \bibnamefont
  {Johnson}},\ }\bibfield  {title} {\bibinfo {title} {Efficient computation of
  the {{3D Green}}'s function for the {{Helmholtz}} operator for a linear array
  of point sources using the {{Ewald}} method},\ }\href
  {https://doi.org/10.1016/j.jcp.2006.09.013} {\bibfield  {journal} {\bibinfo
  {journal} {Journal of Computational Physics}\ }\textbf {\bibinfo {volume}
  {223}},\ \bibinfo {pages} {250} (\bibinfo {year} {2007})}\BibitemShut
  {NoStop}%
\bibitem [{\citenamefont {Chin}\ \emph {et~al.}(1994)\citenamefont {Chin},
  \citenamefont {Nicorovici},\ and\ \citenamefont {McPhedran}}]{chin1994}%
  \BibitemOpen
  \bibfield  {author} {\bibinfo {author} {\bibfnamefont {S.~K.}\ \bibnamefont
  {Chin}}, \bibinfo {author} {\bibfnamefont {N.~A.}\ \bibnamefont
  {Nicorovici}},\ and\ \bibinfo {author} {\bibfnamefont {R.~C.}\ \bibnamefont
  {McPhedran}},\ }\bibfield  {title} {\bibinfo {title} {Green's function and
  lattice sums for electromagnetic scattering by a square array of cylinders},\
  }\href {https://doi.org/10.1103/PhysRevE.49.4590} {\bibfield  {journal}
  {\bibinfo  {journal} {Physical Review E}\ }\textbf {\bibinfo {volume} {49}},\
  \bibinfo {pages} {4590} (\bibinfo {year} {1994})}\BibitemShut {NoStop}%
\bibitem [{\citenamefont {Craeye}\ and\ \citenamefont
  {Capolino}(2006)}]{craeye2006}%
  \BibitemOpen
  \bibfield  {author} {\bibinfo {author} {\bibfnamefont {C.}~\bibnamefont
  {Craeye}}\ and\ \bibinfo {author} {\bibfnamefont {F.}~\bibnamefont
  {Capolino}},\ }\bibfield  {title} {\bibinfo {title} {Accelerated computation
  of the free space {{Green}}'s function of semi-infinite phased arrays of
  dipoles},\ }\href {https://doi.org/10.1109/TAP.2006.869945} {\bibfield
  {journal} {\bibinfo  {journal} {IEEE Transactions on Antennas and
  Propagation}\ }\textbf {\bibinfo {volume} {54}},\ \bibinfo {pages} {1037}
  (\bibinfo {year} {2006})}\BibitemShut {NoStop}%
\bibitem [{\citenamefont {Belov}\ and\ \citenamefont
  {Simovski}(2006)}]{belov2006}%
  \BibitemOpen
  \bibfield  {author} {\bibinfo {author} {\bibfnamefont {P.~A.}\ \bibnamefont
  {Belov}}\ and\ \bibinfo {author} {\bibfnamefont {C.~R.}\ \bibnamefont
  {Simovski}},\ }\bibfield  {title} {\bibinfo {title} {Boundary conditions for
  interfaces of electromagnetic crystals and the generalized {{Ewald-Oseen}}
  extinction principle},\ }\href {https://doi.org/10.1103/PhysRevB.73.045102}
  {\bibfield  {journal} {\bibinfo  {journal} {Physical Review B}\ }\textbf
  {\bibinfo {volume} {73}},\ \bibinfo {pages} {045102} (\bibinfo {year}
  {2006})}\BibitemShut {NoStop}%
\bibitem [{\citenamefont {Dienstfrey}\ \emph {et~al.}(2001)\citenamefont
  {Dienstfrey}, \citenamefont {Hang},\ and\ \citenamefont
  {Huang}}]{dienstfrey2001}%
  \BibitemOpen
  \bibfield  {author} {\bibinfo {author} {\bibfnamefont {A.}~\bibnamefont
  {Dienstfrey}}, \bibinfo {author} {\bibfnamefont {F.}~\bibnamefont {Hang}},\
  and\ \bibinfo {author} {\bibfnamefont {J.}~\bibnamefont {Huang}},\ }\bibfield
   {title} {\bibinfo {title} {Lattice sums and the two-dimensional, periodic
  {{Green}}'s function for the {{Helmholtz}} equation},\ }\href
  {https://doi.org/10.1098/rspa.2000.0656} {\bibfield  {journal} {\bibinfo
  {journal} {Proceedings of the Royal Society of London. Series A:
  Mathematical, Physical and Engineering Sciences}\ }\textbf {\bibinfo {volume}
  {457}},\ \bibinfo {pages} {67} (\bibinfo {year} {2001})}\BibitemShut
  {NoStop}%
\bibitem [{\citenamefont {Jandieri}\ \emph {et~al.}(2019)\citenamefont
  {Jandieri}, \citenamefont {Baccarelli}, \citenamefont {Valerio},\ and\
  \citenamefont {Schettini}}]{jandieri2019}%
  \BibitemOpen
  \bibfield  {author} {\bibinfo {author} {\bibfnamefont {V.}~\bibnamefont
  {Jandieri}}, \bibinfo {author} {\bibfnamefont {P.}~\bibnamefont
  {Baccarelli}}, \bibinfo {author} {\bibfnamefont {G.}~\bibnamefont
  {Valerio}},\ and\ \bibinfo {author} {\bibfnamefont {G.}~\bibnamefont
  {Schettini}},\ }\bibfield  {title} {\bibinfo {title} {1-{{D Periodic Lattice
  Sums}} for {{Complex}} and {{Leaky Waves}} in 2-{{D Structures Using Higher
  Order Ewald Formulation}}},\ }\href
  {https://doi.org/10.1109/TAP.2019.2894280} {\bibfield  {journal} {\bibinfo
  {journal} {IEEE Transactions on Antennas and Propagation}\ }\textbf {\bibinfo
  {volume} {67}},\ \bibinfo {pages} {2364} (\bibinfo {year}
  {2019})}\BibitemShut {NoStop}%
\bibitem [{\citenamefont {McPhedran}\ \emph {et~al.}(2000)\citenamefont
  {McPhedran}, \citenamefont {Nicorovici}, \citenamefont {Botten},\ and\
  \citenamefont {Grubits}}]{mcphedran2000}%
  \BibitemOpen
  \bibfield  {author} {\bibinfo {author} {\bibfnamefont {R.~C.}\ \bibnamefont
  {McPhedran}}, \bibinfo {author} {\bibfnamefont {N.~A.}\ \bibnamefont
  {Nicorovici}}, \bibinfo {author} {\bibfnamefont {L.~C.}\ \bibnamefont
  {Botten}},\ and\ \bibinfo {author} {\bibfnamefont {K.~A.}\ \bibnamefont
  {Grubits}},\ }\bibfield  {title} {\bibinfo {title} {Lattice sums for gratings
  and arrays},\ }\href {https://doi.org/10.1063/1.1310361} {\bibfield
  {journal} {\bibinfo  {journal} {Journal of Mathematical Physics}\ }\textbf
  {\bibinfo {volume} {41}},\ \bibinfo {pages} {7808} (\bibinfo {year}
  {2000})}\BibitemShut {NoStop}%
\bibitem [{\citenamefont {Moroz}(2002)}]{moroz2002}%
  \BibitemOpen
  \bibfield  {author} {\bibinfo {author} {\bibfnamefont {A.}~\bibnamefont
  {Moroz}},\ }\bibfield  {title} {\bibinfo {title} {On the {{Computation}} of
  the {{Free-Space Doubly-Periodic Green}}'s {{Function}} of the
  {{Three-Dimensional Helmholtz Equation}}},\ }\href
  {https://doi.org/10.1163/156939302X00372} {\bibfield  {journal} {\bibinfo
  {journal} {Journal of Electromagnetic Waves and Applications}\ }\textbf
  {\bibinfo {volume} {16}},\ \bibinfo {pages} {457} (\bibinfo {year}
  {2002})}\BibitemShut {NoStop}%
\bibitem [{\citenamefont {Su}\ \emph {et~al.}(2013)\citenamefont {Su},
  \citenamefont {Wu}, \citenamefont {Tang}, \citenamefont {Chen}, \citenamefont
  {Cheng},\ and\ \citenamefont {Jiang}}]{su2013}%
  \BibitemOpen
  \bibfield  {author} {\bibinfo {author} {\bibfnamefont {B.}~\bibnamefont
  {Su}}, \bibinfo {author} {\bibfnamefont {Y.}~\bibnamefont {Wu}}, \bibinfo
  {author} {\bibfnamefont {Y.}~\bibnamefont {Tang}}, \bibinfo {author}
  {\bibfnamefont {Y.}~\bibnamefont {Chen}}, \bibinfo {author} {\bibfnamefont
  {W.}~\bibnamefont {Cheng}},\ and\ \bibinfo {author} {\bibfnamefont
  {L.}~\bibnamefont {Jiang}},\ }\bibfield  {title} {\bibinfo {title}
  {Free-{{Standing 1D Assemblies}} of {{Plasmonic Nanoparticles}}},\ }\href
  {https://doi.org/10.1002/adma.201301003} {\bibfield  {journal} {\bibinfo
  {journal} {Advanced Materials}\ }\textbf {\bibinfo {volume} {25}},\ \bibinfo
  {pages} {3968} (\bibinfo {year} {2013})}\BibitemShut {NoStop}%
\bibitem [{\citenamefont {Kruk}\ \emph {et~al.}(2017)\citenamefont {Kruk},
  \citenamefont {Slobozhanyuk}, \citenamefont {Denkova}, \citenamefont
  {Poddubny}, \citenamefont {Kravchenko}, \citenamefont {Miroshnichenko},
  \citenamefont {Neshev},\ and\ \citenamefont {Kivshar}}]{kruk2017}%
  \BibitemOpen
  \bibfield  {author} {\bibinfo {author} {\bibfnamefont {S.}~\bibnamefont
  {Kruk}}, \bibinfo {author} {\bibfnamefont {A.}~\bibnamefont {Slobozhanyuk}},
  \bibinfo {author} {\bibfnamefont {D.}~\bibnamefont {Denkova}}, \bibinfo
  {author} {\bibfnamefont {A.}~\bibnamefont {Poddubny}}, \bibinfo {author}
  {\bibfnamefont {I.}~\bibnamefont {Kravchenko}}, \bibinfo {author}
  {\bibfnamefont {A.}~\bibnamefont {Miroshnichenko}}, \bibinfo {author}
  {\bibfnamefont {D.}~\bibnamefont {Neshev}},\ and\ \bibinfo {author}
  {\bibfnamefont {Y.}~\bibnamefont {Kivshar}},\ }\bibfield  {title} {\bibinfo
  {title} {Edge {{States}} and {{Topological Phase Transitions}} in {{Chains}}
  of {{Dielectric Nanoparticles}}},\ }\href
  {https://doi.org/10.1002/smll.201603190} {\bibfield  {journal} {\bibinfo
  {journal} {Small}\ }\textbf {\bibinfo {volume} {13}},\ \bibinfo {pages}
  {1603190} (\bibinfo {year} {2017})}\BibitemShut {NoStop}%
\bibitem [{\citenamefont {Chen}\ and\ \citenamefont {Rosi}(2010)}]{chen2010}%
  \BibitemOpen
  \bibfield  {author} {\bibinfo {author} {\bibfnamefont {C.-L.}\ \bibnamefont
  {Chen}}\ and\ \bibinfo {author} {\bibfnamefont {N.~L.}\ \bibnamefont
  {Rosi}},\ }\bibfield  {title} {\bibinfo {title} {Preparation of {{Unique}}
  1-{{D Nanoparticle Superstructures}} and {{Tailoring}} their {{Structural
  Features}}},\ }\href {https://doi.org/10.1021/ja102000g} {\bibfield
  {journal} {\bibinfo  {journal} {Journal of the American Chemical Society}\
  }\textbf {\bibinfo {volume} {132}},\ \bibinfo {pages} {6902} (\bibinfo {year}
  {2010})}\BibitemShut {NoStop}%
\bibitem [{\citenamefont {Rechtsman}\ \emph {et~al.}(2013)\citenamefont
  {Rechtsman}, \citenamefont {Zeuner}, \citenamefont {Plotnik}, \citenamefont
  {Lumer}, \citenamefont {Podolsky}, \citenamefont {Dreisow}, \citenamefont
  {Nolte}, \citenamefont {Segev},\ and\ \citenamefont
  {Szameit}}]{rechtsman2013}%
  \BibitemOpen
  \bibfield  {author} {\bibinfo {author} {\bibfnamefont {M.~C.}\ \bibnamefont
  {Rechtsman}}, \bibinfo {author} {\bibfnamefont {J.~M.}\ \bibnamefont
  {Zeuner}}, \bibinfo {author} {\bibfnamefont {Y.}~\bibnamefont {Plotnik}},
  \bibinfo {author} {\bibfnamefont {Y.}~\bibnamefont {Lumer}}, \bibinfo
  {author} {\bibfnamefont {D.}~\bibnamefont {Podolsky}}, \bibinfo {author}
  {\bibfnamefont {F.}~\bibnamefont {Dreisow}}, \bibinfo {author} {\bibfnamefont
  {S.}~\bibnamefont {Nolte}}, \bibinfo {author} {\bibfnamefont
  {M.}~\bibnamefont {Segev}},\ and\ \bibinfo {author} {\bibfnamefont
  {A.}~\bibnamefont {Szameit}},\ }\bibfield  {title} {\bibinfo {title}
  {Photonic {{Floquet}} topological insulators},\ }\href
  {https://doi.org/10.1038/nature12066} {\bibfield  {journal} {\bibinfo
  {journal} {Nature}\ }\textbf {\bibinfo {volume} {496}},\ \bibinfo {pages}
  {196} (\bibinfo {year} {2013})}\BibitemShut {NoStop}%
\bibitem [{\citenamefont {Theobald}\ \emph {et~al.}(2021)\citenamefont
  {Theobald}, \citenamefont {Beutel}, \citenamefont {Borgmann}, \citenamefont
  {Mescher}, \citenamefont {Gomard}, \citenamefont {Rockstuhl},\ and\
  \citenamefont {Lemmer}}]{theobald2021}%
  \BibitemOpen
  \bibfield  {author} {\bibinfo {author} {\bibfnamefont {D.}~\bibnamefont
  {Theobald}}, \bibinfo {author} {\bibfnamefont {D.}~\bibnamefont {Beutel}},
  \bibinfo {author} {\bibfnamefont {L.}~\bibnamefont {Borgmann}}, \bibinfo
  {author} {\bibfnamefont {H.}~\bibnamefont {Mescher}}, \bibinfo {author}
  {\bibfnamefont {G.}~\bibnamefont {Gomard}}, \bibinfo {author} {\bibfnamefont
  {C.}~\bibnamefont {Rockstuhl}},\ and\ \bibinfo {author} {\bibfnamefont
  {U.}~\bibnamefont {Lemmer}},\ }\bibfield  {title} {\bibinfo {title}
  {Simulation of light scattering in large, disordered nanostructures using a
  periodic {{T-matrix}} method},\ }\href
  {https://doi.org/10.1016/j.jqsrt.2021.107802} {\bibfield  {journal} {\bibinfo
   {journal} {Journal of Quantitative Spectroscopy and Radiative Transfer}\
  }\textbf {\bibinfo {volume} {272}},\ \bibinfo {pages} {107802} (\bibinfo
  {year} {2021})}\BibitemShut {NoStop}%
\bibitem [{\citenamefont {Moritake}\ \emph {et~al.}(2022)\citenamefont
  {Moritake}, \citenamefont {Ono},\ and\ \citenamefont
  {Notomi}}]{moritake2022}%
  \BibitemOpen
  \bibfield  {author} {\bibinfo {author} {\bibfnamefont {Y.}~\bibnamefont
  {Moritake}}, \bibinfo {author} {\bibfnamefont {M.}~\bibnamefont {Ono}},\ and\
  \bibinfo {author} {\bibfnamefont {M.}~\bibnamefont {Notomi}},\ }\bibfield
  {title} {\bibinfo {title} {Far-field optical imaging of topological edge
  states in zigzag plasmonic chains},\ }\href
  {https://doi.org/10.1515/nanoph-2021-0648} {\bibfield  {journal} {\bibinfo
  {journal} {Nanophotonics}\ }\textbf {\bibinfo {volume} {11}},\ \bibinfo
  {pages} {2183} (\bibinfo {year} {2022})}\BibitemShut {NoStop}%
\bibitem [{\citenamefont {Koshelev}\ \emph {et~al.}(2018)\citenamefont
  {Koshelev}, \citenamefont {Lepeshov}, \citenamefont {Liu}, \citenamefont
  {Bogdanov},\ and\ \citenamefont {Kivshar}}]{koshelev2018}%
  \BibitemOpen
  \bibfield  {author} {\bibinfo {author} {\bibfnamefont {K.}~\bibnamefont
  {Koshelev}}, \bibinfo {author} {\bibfnamefont {S.}~\bibnamefont {Lepeshov}},
  \bibinfo {author} {\bibfnamefont {M.}~\bibnamefont {Liu}}, \bibinfo {author}
  {\bibfnamefont {A.}~\bibnamefont {Bogdanov}},\ and\ \bibinfo {author}
  {\bibfnamefont {Y.}~\bibnamefont {Kivshar}},\ }\bibfield  {title} {\bibinfo
  {title} {Asymmetric {{Metasurfaces}} with {{High- Q Resonances Governed}} by
  {{Bound States}} in the {{Continuum}}},\ }\href
  {https://doi.org/10.1103/PhysRevLett.121.193903} {\bibfield  {journal}
  {\bibinfo  {journal} {Physical Review Letters}\ }\textbf {\bibinfo {volume}
  {121}},\ \bibinfo {pages} {193903} (\bibinfo {year} {2018})}\BibitemShut
  {NoStop}%
\bibitem [{\citenamefont {Sun}\ \emph {et~al.}(2017)\citenamefont {Sun},
  \citenamefont {Dong}, \citenamefont {Si},\ and\ \citenamefont
  {Deng}}]{sun2017}%
  \BibitemOpen
  \bibfield  {author} {\bibinfo {author} {\bibfnamefont {C.}~\bibnamefont
  {Sun}}, \bibinfo {author} {\bibfnamefont {Z.}~\bibnamefont {Dong}}, \bibinfo
  {author} {\bibfnamefont {J.}~\bibnamefont {Si}},\ and\ \bibinfo {author}
  {\bibfnamefont {X.}~\bibnamefont {Deng}},\ }\bibfield  {title} {\bibinfo
  {title} {Independently tunable dual-band plasmonically induced transparency
  based on hybrid metal-graphene metamaterials at mid-infrared frequencies},\
  }\href {https://doi.org/10.1364/OE.25.001242} {\bibfield  {journal} {\bibinfo
   {journal} {Optics Express}\ }\textbf {\bibinfo {volume} {25}},\ \bibinfo
  {pages} {1242} (\bibinfo {year} {2017})}\BibitemShut {NoStop}%
\bibitem [{\citenamefont {Wang}\ \emph {et~al.}(2020)\citenamefont {Wang},
  \citenamefont {Zheng}, \citenamefont {Chen}, \citenamefont {Huang},
  \citenamefont {Kartashov}, \citenamefont {Torner}, \citenamefont {Konotop},\
  and\ \citenamefont {Ye}}]{wang2020}%
  \BibitemOpen
  \bibfield  {author} {\bibinfo {author} {\bibfnamefont {P.}~\bibnamefont
  {Wang}}, \bibinfo {author} {\bibfnamefont {Y.}~\bibnamefont {Zheng}},
  \bibinfo {author} {\bibfnamefont {X.}~\bibnamefont {Chen}}, \bibinfo {author}
  {\bibfnamefont {C.}~\bibnamefont {Huang}}, \bibinfo {author} {\bibfnamefont
  {Y.~V.}\ \bibnamefont {Kartashov}}, \bibinfo {author} {\bibfnamefont
  {L.}~\bibnamefont {Torner}}, \bibinfo {author} {\bibfnamefont {V.~V.}\
  \bibnamefont {Konotop}},\ and\ \bibinfo {author} {\bibfnamefont
  {F.}~\bibnamefont {Ye}},\ }\bibfield  {title} {\bibinfo {title} {Localization
  and delocalization of light in photonic moir\'e lattices},\ }\href
  {https://doi.org/10.1038/s41586-019-1851-6} {\bibfield  {journal} {\bibinfo
  {journal} {Nature}\ }\textbf {\bibinfo {volume} {577}},\ \bibinfo {pages}
  {42} (\bibinfo {year} {2020})}\BibitemShut {NoStop}%
\bibitem [{\citenamefont {Ne{\v c}ada}\ and\ \citenamefont
  {T{\"o}rm{\"a}}(2021)}]{necada2021}%
  \BibitemOpen
  \bibfield  {author} {\bibinfo {author} {\bibfnamefont {M.}~\bibnamefont
  {Ne{\v c}ada}}\ and\ \bibinfo {author} {\bibfnamefont {P.}~\bibnamefont
  {T{\"o}rm{\"a}}},\ }\bibfield  {title} {\bibinfo {title}
  {Multiple-{{Scattering}} ${{T}}$-{{Matrix Simulations}} for
  {{Nanophotonics}}: {{Symmetries}} and {{Periodic Lattices}}},\ }\href
  {https://doi.org/10.4208/cicp.OA-2020-0136} {\bibfield  {journal} {\bibinfo
  {journal} {Communications in Computational Physics}\ }\textbf {\bibinfo
  {volume} {30}},\ \bibinfo {pages} {357} (\bibinfo {year} {2021})}\BibitemShut
  {NoStop}%
\bibitem [{\citenamefont {Solbrig}(1982)}]{solbrig1982}%
  \BibitemOpen
  \bibfield  {author} {\bibinfo {author} {\bibfnamefont {H.}~\bibnamefont
  {Solbrig}},\ }\bibfield  {title} {\bibinfo {title} {On the {{Ewald Summation
  Technique}} for {{2D Lattices}}},\ }\href
  {https://doi.org/10.1002/pssa.2210720120} {\bibfield  {journal} {\bibinfo
  {journal} {physica status solidi (a)}\ }\textbf {\bibinfo {volume} {72}},\
  \bibinfo {pages} {199} (\bibinfo {year} {1982})}\BibitemShut {NoStop}%
\bibitem [{\citenamefont {Kambe}(1967)}]{kambe1967}%
  \BibitemOpen
  \bibfield  {author} {\bibinfo {author} {\bibfnamefont {K.}~\bibnamefont
  {Kambe}},\ }\bibfield  {title} {\bibinfo {title} {Theory of {{Low}}-{{Energy
  Electron Diffraction}} ({{I}}. {{Application}} of the {{Cellular Method}} to
  {{Monatomic Layers}})},\ }\href {https://doi.org/10.1515/zna-1967-0305}
  {\bibfield  {journal} {\bibinfo  {journal} {Zeitschrift f\"ur Naturforschung
  A}\ }\textbf {\bibinfo {volume} {22}},\ \bibinfo {pages} {322} (\bibinfo
  {year} {1967})}\BibitemShut {NoStop}%
\bibitem [{\citenamefont {Kambe}(1968)}]{kambe1968}%
  \BibitemOpen
  \bibfield  {author} {\bibinfo {author} {\bibfnamefont {K.}~\bibnamefont
  {Kambe}},\ }\bibfield  {title} {\bibinfo {title} {Theory of {{Low}}-{{Energy
  Electron Diffraction}} ({{II}}. {{Cellular Method}} for {{Complex
  Monolayers}} and {{Multilayers}})},\ }\href
  {https://doi.org/10.1515/zna-1968-0908} {\bibfield  {journal} {\bibinfo
  {journal} {Zeitschrift f\"ur Naturforschung A}\ }\textbf {\bibinfo {volume}
  {23}},\ \bibinfo {pages} {1280} (\bibinfo {year} {1968})}\BibitemShut
  {NoStop}%
\bibitem [{\citenamefont {Waterman}(1965)}]{waterman1965}%
  \BibitemOpen
  \bibfield  {author} {\bibinfo {author} {\bibfnamefont {P.~C.}\ \bibnamefont
  {Waterman}},\ }\bibfield  {title} {\bibinfo {title} {Matrix formulation of
  electromagnetic scattering},\ }\href {https://doi.org/10.1109/proc.1965.4058}
  {\bibfield  {journal} {\bibinfo  {journal} {Proceedings of the IEEE}\
  }\textbf {\bibinfo {volume} {53}},\ \bibinfo {pages} {805} (\bibinfo {year}
  {1965})}\BibitemShut {NoStop}%
\bibitem [{\citenamefont {Mishchenko}(2020)}]{mishchenko2020}%
  \BibitemOpen
  \bibfield  {author} {\bibinfo {author} {\bibfnamefont {M.~I.}\ \bibnamefont
  {Mishchenko}},\ }\bibfield  {title} {\bibinfo {title} {Comprehensive thematic
  {{T-matrix}} reference database: A 2017\textendash 2019 update},\ }\href
  {https://doi.org/10.1016/j.jqsrt.2019.106692} {\bibfield  {journal} {\bibinfo
   {journal} {Journal of Quantitative Spectroscopy and Radiative Transfer}\
  }\textbf {\bibinfo {volume} {242}},\ \bibinfo {pages} {106692} (\bibinfo
  {year} {2020})}\BibitemShut {NoStop}%
\bibitem [{\citenamefont {Eyert}(2012)}]{eyert2012}%
  \BibitemOpen
  \bibfield  {author} {\bibinfo {author} {\bibfnamefont {V.}~\bibnamefont
  {Eyert}},\ }\href@noop {} {\emph {\bibinfo {title} {The Augmented Spherical
  Wave Method: A Comprehensive Treatment}}},\ \bibinfo {edition} {second
  edition}\ ed.,\ \bibinfo {series} {Lecture Notes in Physics}\ No.\ \bibinfo
  {number} {849}\ (\bibinfo  {publisher} {{Springer}},\ \bibinfo {address}
  {{Heidelberg}},\ \bibinfo {year} {2012})\BibitemShut {NoStop}%
\bibitem [{\citenamefont {Grad{\v s}tejn}\ and\ \citenamefont {Ry{\v
  z}ik}(2014)}]{gradstejn2014}%
  \BibitemOpen
  \bibfield  {author} {\bibinfo {author} {\bibfnamefont {I.~S.}\ \bibnamefont
  {Grad{\v s}tejn}}\ and\ \bibinfo {author} {\bibfnamefont {I.~M.}\
  \bibnamefont {Ry{\v z}ik}},\ }\href@noop {} {\emph {\bibinfo {title} {Table
  of {{Integrals}}, {{Series}}, and {{Products}}}}},\ edited by\ \bibinfo
  {editor} {\bibfnamefont {D.}~\bibnamefont {Zwillinger}}\ (\bibinfo
  {publisher} {{Elsevier Science}},\ \bibinfo {year} {2014})\BibitemShut
  {NoStop}%
\bibitem [{\citenamefont {Morse}\ and\ \citenamefont
  {Feshbach}(1953)}]{morse1953}%
  \BibitemOpen
  \bibfield  {author} {\bibinfo {author} {\bibfnamefont {P.~M.}\ \bibnamefont
  {Morse}}\ and\ \bibinfo {author} {\bibfnamefont {H.}~\bibnamefont
  {Feshbach}},\ }\href@noop {} {\emph {\bibinfo {title} {Methods of Theoretical
  Physics}}}\ (\bibinfo  {publisher} {{McGraw-Hill}},\ \bibinfo {address} {{New
  York}},\ \bibinfo {year} {1953})\BibitemShut {NoStop}%
\bibitem [{\citenamefont {Magnus}\ \emph {et~al.}(1966)\citenamefont {Magnus},
  \citenamefont {Oberhettinger},\ and\ \citenamefont {Soni}}]{magnus1966}%
  \BibitemOpen
  \bibfield  {author} {\bibinfo {author} {\bibfnamefont {W.}~\bibnamefont
  {Magnus}}, \bibinfo {author} {\bibfnamefont {F.}~\bibnamefont
  {Oberhettinger}},\ and\ \bibinfo {author} {\bibfnamefont {R.~P.}\
  \bibnamefont {Soni}},\ }\href@noop {} {\emph {\bibinfo {title} {Formulas and
  Theorems for the Special Functions of Mathematical Physics}}},\ \bibinfo
  {edition} {3rd}\ ed.,\ \bibinfo {series} {Die {{Grundlehren}} Der
  Mathematischen {{Wissenschaften}} in {{Einzeldarstellungen}}}, Vol.~\bibinfo
  {volume} {52}\ (\bibinfo  {publisher} {{Springer}},\ \bibinfo {year}
  {1966})\BibitemShut {NoStop}%
\end{thebibliography}%

\appendix

\section{Spherical harmonics and associated Legendre polynomials}
\label{app:sphharm}
The spherical harmonics we use are defined by
\begin{equation}\label{eq:sphharm}
    Y_{l m}(\theta, \phi)
    =
    \underset{N_{l m}}{\underbrace{\sqrt{\frac{2l - 1}{4\pi}\frac{(l - m)!}{(l + m)!}}}}
    P_l^m (\cos\theta)
    \ee^{\ii m \phi}
    \,,
\end{equation}
where $P_l^m(x)$ are the Legendre polynomials
\begin{align} \label{eq:legendrep:start}
    P_l^m (x)
    =&
    (-1)^m
    \left(1 - x^2\right)^{\frac{m}{2}}
    \frac{\dd^m}{\dd x^m} P_l(x) \\
    =&
    \frac{(-1)^m}{2^l l!}
    \left(1 - x^2\right)^{\frac{m}{2}}
    \frac{\dd^{l + m}}{\dd x^{l + m}}
    \left(x^2 - 1\right)^l \label{eq:legendrep}
    \,,
\end{align}
where \cref{eq:legendrep:start} defines the associated Legendre polynomials, in principle, only for $m \geq 0$. After using using Rodrigues' formula for the Legendre polynomials,yy to arrive at \cref{eq:legendrep}, the expression can be used for all $|m| \leq l$.

To derive the closed form expression for the associated Legendre polynomials in the main text, we begin with
\begin{align}
    P_l^{|m|}(\cos\theta_{\bm{r}}) = 
    P_l^{|m|} \left(\frac{z}{\sqrt{\rho^2 + z^2}}\right) \notag
    \\
    = \sum_{j = 0}^{\lfloor\frac{l - |m|}{2}\rfloor}
    \frac{(-1)^{j + |m|} (l - 2j)!}{2^l (l - |m| - 2j)!} \binom{l}{j} \binom{2l - 2j}{l}
    \frac{z^{l - |m| - 2j} \rho^{|m|}}{\sqrt{\rho^2 + z^2}^{l - 2j}}
\end{align}
where $\bm{r} = (x, y, z)^T$ and $\rho = \sqrt{x^2 + y^2}$,
for the associated Legendre polynomials, which can be derived by evaluating \cref{eq:legendrep:start}
and using the closed expression
\begin{equation}\label{eq:rodrigues}
    P_l(x)
    = \frac{1}{2^l}
    \sum_{j = 0}^{\lfloor \frac{l}{2} \rfloor}
    (-1)^j
    \binom{l}{j}
    \binom{2l - 2j}{l} x^{l - 2j}
\end{equation}
for the Legendre polynomials~\cite[Eq. 8.911 1.]{gradstejn2014}.
We expand $\sqrt{\rho^2 + z^2}^{2j}$ to arrive at
\begin{align}
    P_l^m& \left(\frac{z}{\sqrt{\rho^2 + z^2}}\right)
    \notag\\
    =& \frac{(-1)^{|m|}\rho^{|m|}}{2^l \sqrt{\rho^2 + z^2}^l}
    \sum_{j = 0}^{\lfloor\frac{l - |m|}{2}\rfloor}
    \frac{(-1)^j (l - 2j)!}{(l - |m| - 2j)!} \binom{l}{j} \binom{2l - 2j}{l}
    \notag \\
    &\cdot
    \sum_{s = 0}^j
    \binom{j}{s}
    \rho^{2s} z^{l - |m| - 2s}\,.
\end{align}
Now, we can rearrange the series to
\begin{align} \label{eq:legendrep:intermediate}
    P_l^m& \left(\frac{z}{\sqrt{\rho^2 + z^2}}\right) \notag\\
    =& \frac{(-1)^{|m|}\rho^{|m|}}{2^l \sqrt{\rho^2 + z^2}^l}
    \sum_{s = 0}^{\lfloor\frac{l - |m|}{2}\rfloor}
    \frac{\rho^{2s} z^{l - |m| - 2s}}{s!} \notag \\
    &\cdot
    \sum_{j = s}^{\lfloor\frac{l - |m|}{2}\rfloor}
    \frac{(-1)^j (2l - 2j)!}{(l - |m| - 2j)! (l - j)! (j - s)!}
\end{align}
where the last sum fulfills 
\begin{align} \label{eq:legendrep:recursion}
    f(l, m, s)
    &=
    \sum_{j = s}^{\lfloor\frac{l - |m|}{2}\rfloor}
    \frac{(-1)^j (2l - 2j)!}{(l - |m| - 2j)! (l - j)! (j - s)!} \\
    &=
    \frac{(-1)^s(l + m)! 2^{l - m - 2s}}
    {(l - m - 2s)! (s + m)!}
    \,,
\end{align}
which can be shown by using the recursion formula
\begin{equation}
    f(l + 1, m, s) = 2 (f(l, m ,s) + (l + m) f(l, m - 1, s)
\end{equation}
and the initial condition
\begin{equation}
    f(l, -l, s) = \delta_{ls} (-1)^l\,.
\end{equation}
Thus, combining \cref{eq:legendrep:intermediate} and \cref{eq:legendrep:recursion}, we arrive at
\begin{align}\label{eq:legendrep:closedform}
    P_l^m& \left(\frac{z}{\sqrt{\rho^2 + z^2}}\right) \notag\\
    =& \frac{(-1)^{\frac{|m| + m}{2}}}{\sqrt{\rho^2 + z^2}^l}
    \sum_{s = 0}^{\lfloor\frac{l - |m|}{2}\rfloor}
    \rho^{2s + |m|} z^{l - |m| - 2s} \notag \\
    &\cdot
    \frac{(-1)^s (l + m)!}{2^{2s + |m|}(l - |m| - 2s)!(s + |m|)! s!}
\end{align}
as our final expression for the associated Legendre polynomials, which we have generalized
to negative values of $m$ with
\begin{equation}
    P_l^{-m}(x) = (-1)^m \frac{(l - m)!}{(l + m)!} P_l^m(x)\,.
\end{equation}

\section{Plane wave expansion}
\label{app:pwexp}
We use the expressions~\cite{morse1953}
\begin{align}
    \label{eq:pwexp:cw}
    \mathrm{e}^{-\mathrm{i} \bm{kr}}
    = \sum_{l = -\infty}^\infty
    (-\mathrm{i})^{|l|} J_{|l|}(kr)
    \mathrm{e}^{\mathrm{i} l (\phi_{\bm{k}} - \phi_{\bm{r}})}
\end{align}
if $\bm{k}, \bm{r} \in \mathbb{R}^2$ and
\begin{align}
    \label{eq:pwexp:sw}
    \mathrm{e}^{-\mathrm{i} \bm{kr}}
    =& 4 \pi \sum_{l = 0}^\infty \sum_{m = -l}^l
    (-\mathrm{i})^{l} j_{l}(kr)
    Y_{lm}(\theta_{\bm{k}}, \phi_{\bm{k}})
    Y_{lm}^\ast(\theta_{\bm{r}}, \phi_{\bm{r}})
\end{align}
if $\bm{k}, \bm{r} \in \mathbb{R}^3$ to expand the plane waves using cylindrical and spherical coordinates.

\section{Real and reciprocal space integral}
\label{app:realrecint}
The integral
\begin{align}
    I_n(x, \alpha) = \int\limits_\alpha^\infty \mathrm d t \,t^n
    \mathrm e^{-\frac{z^2t^2}{2} + \frac{1}{2t^2}}
\end{align}
used for the real space part of the sum
fulfils the  recursion relation~\cite{kambe1967}
\begin{align}\label{eq:realint:recursion}
    I_n(z, \alpha)
    =
    (n + 3) I_{n + 2}(z, \alpha)
    - z^2 I_{n + 4}(z, \alpha)
    \notag \\
    + \alpha^{n + 3} \mathrm e^{-\frac{z^2\alpha^2}{2} + \frac{1}{2\alpha^2}}
    \,,
\end{align}
which can also be rearranged for increasing values of $n$ instead of decreasing values. As initial values, two integrals have to be known for odd and even values of $n$, so in total 4 integrals. We evaluate the integrals for $n \in \{-3, -2, -1, 0\}$ directly. For $n = -2$ and $n = 0$, we construct the new integral
\begin{align}
    z I_0(z, \alpha) \pm \mathrm i I_{-2}(z, \alpha)
    = \int\limits_\alpha^\infty \mathrm d t \left(z \pm \frac{\mathrm i}{t^2}\right)
    \mathrm e^{-\frac{(zt \mp \frac{\mathrm i}{t})^2}{2} \mp \mathrm i z}
    \notag \\
    = \mathrm{e}^{\mp\mathrm{i}z} \sqrt{2}
    \int\limits_{\frac{1}{\sqrt{2}}(\alpha z \mp \frac{\mathrm i}{\alpha})}^\infty
    \mathrm d u\, \mathrm e^{-u^2}
    =
    \sqrt{\frac{\pi}{2}} \mathrm{e}^{\mp \mathrm{i} z}
    \erfc\left(\frac{\alpha z \mp \frac{\mathrm i}{\alpha}}{\sqrt{2}}\right)\,.
\end{align}
With this result, the required initial integrals for the recursion over even numbers is
\begin{widetext}
\begin{align}
    I_0(z, \alpha) = \frac{\sqrt{\pi}}{2\sqrt{2}z}
    \left(
    \mathrm{e}^{-\mathrm i z} \erfc\left(\frac{\alpha z - \frac{\mathrm i}{\alpha}}{\sqrt{2}}\right)
    + \mathrm{e}^{\mathrm i z} \erfc\left(\frac{\alpha z + \frac{\mathrm i}{\alpha}}{\sqrt{2}}\right)
    \right) \\
    I_{-2}(z, \alpha) = \frac{-\mathrm i\sqrt{\pi}}{2\sqrt{2}z}
    \left(
    \mathrm{e}^{-\mathrm i z} \erfc\left(\frac{\alpha z - \frac{\mathrm i}{\alpha}}{\sqrt{2}}\right)
    - \mathrm{e}^{\mathrm i z} \erfc\left(\frac{\alpha z + \frac{\mathrm i}{\alpha}}{\sqrt{2}}\right)
    \right)\,.
\end{align}
\end{widetext}
In the case $n = -1$, the integral becomes after a substitution $u=\tfrac{z^2t^2}{2}$
\begin{subequations}
\begin{align}
    I_{-1}(z, \alpha) &= \frac{1}{2} \int\limits_{\frac{z^2 t^2}{2}}^\infty \frac{\mathrm d u}{u}
    \mathrm e^{-u} \mathrm e^{\frac{z^2}{4u}} \\
    &= \frac{1}{2} \sum_{n = 0}^\infty \frac{1}{n!} \left(\frac{z^2}{4}\right)^n
    \int\limits_{\frac{z^2 t^2}{2}}^\infty \mathrm d u\,u^{-n-1}
    \mathrm e^{-u} \\
    &= \frac{1}{2} \sum_{n = 0}^\infty \frac{1}{n!} \left(\frac{z^2}{4}\right)^n
    \Gamma\left(-n, \frac{z^2 t^2}{2}\right)\,.
\end{align}
\end{subequations}
This summation converges quite fast and can be truncated for a numerical evaluation. Similarly, we derive
\begin{align}
    I_{-3}(z, \alpha)
    = \sum_{n = 0}^\infty \frac{1}{n!} \left(\frac{z^2}{4}\right)^{n + 1}
    \Gamma\left(-n - 1, \frac{z^2 t^2}{2}\right)\,.
\end{align}
With these four starting values we can use the recursion formula for positive and negative values of $n$.

The reciprocal space integral reads
\begin{equation}
    \int\limits_{-\frac{\gamma^2}{2\eta^2}}^\infty \frac{\dd u}{u}\, u^n
    \ee^{-u+\frac{(\gamma k z)^2}{4u}}
\end{equation}
for $n$ either integer or half integer numbers. This integral can be transformed to the integral $I_l$.
For this, we take the substitution $t = \tfrac{\sqrt{2u}}{k \gamma z}$ resulting in
\begin{align}
    2
    \left( \frac{k^2 \gamma^2 z^2}{2} \right)^n
    \int\limits_{-\frac{\ii}{k z \eta}}^\infty \dd t\, t^{2n - 1}
    \ee^{-\frac{(k \gamma z t)^2}{2} + \frac{1}{2 t^2}} \notag \\
    =
    2
    \left( \frac{k^2 \gamma^2 z^2}{2} \right)^n
    I_{2n - 1}\left(k \gamma z, -\frac{\ii}{k z \eta}\right)
    \,,
\end{align}
which has the exact same form as the real space integral. Therefore, it can be calculated with \cref{eq:realint:recursion} in combination with the previously derived initial values.

\section{Sum manipulations}
In the main text, the following manipulations of the summation indices are used.
\subsection{$d = 3$, $d' = 2$}
\begin{equation}\label{eq:sumswap:sw2d}
\begin{split}
    \sum_{s = 0}^{\left\lfloor \frac{l - |m|}{2} \right\rfloor}
    \sum_{n = 0}^s
    a_{s, n}
    &
    =
    \sum_{s = 0}^{\left\lfloor \frac{l - |m|}{2} \right\rfloor}
    \sum_{w = l - |m| - 2s}^{l - |m| - s}
    a_{s, w - l + |m| + 2s}
    \\
    &
    =
    \sum_{w = 0}^{l - |m|}
    \sum_{s = \left\lceil \frac{l - |m| - n}{2} \right\rceil}^{\min\left(l - |m|, \left\lfloor\frac{l - |m|}{2}\right\rfloor\right)}
    a_{s, w - l + |m| + 2s}
    \\
    &
    =
    \sum_{w = 0}^{l - |m|}
    \sum_{v = w}^{\min(l - |m|, 2w)}
    a_{\frac{l - |m| - 2w + v}{2}, v - w}
    \end{split}
\end{equation}
where $v$ in the last line only takes values with the same parity as $l - |m|$.

\subsection{$d = 3$, $d' = 1$}
\begin{equation}\label{eq:sumswap:sw1d}
    \begin{split}
    \sum_{s = 0}^{\lfloor\frac{l - |m|}{2}\rfloor}
    \sum_{n = 0}^{\left\lfloor\frac{l - |m|}{2}\right\rfloor - s}
    &
    a_{s, n}
    =
    \sum_{s = 0}^{\lfloor\frac{l - |m|}{2}\rfloor}
    \sum_{w = |m| + 2s}^{\left\lfloor\frac{l + |m|}{2}\right\rfloor + s}
    a_{s, w - |m| - 2s}
    \\
    &
    =
    \sum_{w = |m|}^l
    \sum_{s = \max\left(0, n -\lfloor\frac{l + |m|}{2}\rfloor\right)}^{\left\lfloor\frac{w - |m|}{2}\right\rfloor}
    a_{s, w - |m| - 2s}
    \\
    &
    =
    \sum_{w = |m|}^l
    \sum_{v = w}^{\min(2n - |m|, l)}
    a_{w - \frac{v + |m|}{2}, v - w}
\end{split}
\end{equation}
where $v$ in the last line only takes values with the same parity as $|m|$.

\subsection{$d = 2$, $d' = 1$}
\begin{equation}\label{eq:sumswap:cw1d}
    \begin{split}
    \sum_{s = 0}^{|l|}
    \sum_{n = 0}^{\left\lfloor\frac{s}{2}\right\rfloor}
    a_{s, n}
    &
    =
    \sum_{s = 0}^{|l|}
    \sum_{w = |l| - s}^{|l| - \left\lceil\frac{s}{2}\right\rceil}
    a_{s, w + s - |l|}
    \\
    &
    =
    \sum_{w = 0}^{|l|}
    \sum_{s = |l| - w}^{\min(|l|, 2|l| - 2w)}
    a_{s, w + s - |l|}
    \\
    &
    =
    \sum_{w = 0}^{|l|}
    \sum_{v = w}^{\min(2w, |l|)}
    a_{v + |l| - 2w, v - w}
\end{split}
\end{equation}

\section{Whittaker function}

It holds that
\begin{align}
    M_{\frac{1 + |m|}{2} + s, \frac{|m|}{2}}(z)
    =
    \frac{\ee^{\frac{z}{2}} z^\frac{1 - |m|}{2} |m|!}{(|m| + s)!}
    \frac{\dd^s}{\dd z^s}(\ee^{-z} z^{|m|+ s})
\end{align}
for $n \in \mathbb{N}$ and $m \in \mathbb{Z}$~\cite[sec. 7.2.4]{magnus1966}.
With the generalized product rule for derivatives
\begin{align}
    \frac{\dd^s}{\dd z^s}(f(z) g(z))
    = \sum_{n = 0}^s
    \binom{s}{n}
    \left(\frac{\dd^{s - n}}{\dd z^{s - n}}f(z)\right)
    \left(\frac{\dd^n}{\dd z^n}g(z)\right)
    \,,
\end{align}
we obtain the expression
\begin{align}\label{eq:whittaker}
    M_{\frac{1 + |m|}{2} + s, \frac{|m|}{2}}(z)
    =
    \ee^{-\frac{z}{2}} |m|!
    \sum_{n = 0}^s
    \binom{s}{n}
    \frac{z^{\frac{1 + |m|}{2}}(-z)^{s - n}}{(|m| + s - n)!}
    \,.
\end{align}

\section{Simplifications for $\bm r_\perp = 0$}\label{app:simplify}
The following simplifications in the reciprocal space sum can be obtained for a vanishing shift perpendicular to the lattice
\begin{widetext}
\begin{align}
    \label{eq:simplify:s32}
    S_{3, l m n, 2}(k, \beta, 0)
    &= \begin{cases}
        \frac{\sqrt{(2l + 1)(l - m)! (l + m)!}}
        {(-2)^l V_2 k^2}
        \frac{\beta^{l - 2n}}
        {n! \left(\frac{l - m}{2} - n\right) \left(\frac{l + m}{2} - n\right)}
        &
        n \leq \left\lfloor \frac{l - |m|}{2} \right\rfloor
        \text{ and } l - m \text{ even}
        \\
        0 & \text{otherwise}
    \end{cases}
    \\
    \label{eq:simplify:s31}
    S_{3, l m n, 1}(k, \beta, 0)
    &= \begin{cases}
        \frac{(-\ii)^{l + 1} l!}
        {2 V_1 k}
        \sqrt{\frac{2l + 1}{\pi}}
        \frac{\beta^{l - 2n}}
        {n! (l - 2n)!}
        &
        n \leq \left\lfloor \frac{l}{2} \right\rfloor
        \text{ and } m = 0
        \\
        0 & \text{otherwise}
    \end{cases}
    \\
    \label{eq:simplify:s21}
    S_{2, l n, 1}(k, \beta, 0)
    &= \begin{cases}
        \frac{2 (-\ii)^l}
        {\sqrt{\pi} V_1 k}
        \frac{\beta^{|l| - 2n}}
        {4^n n! (|l| - 2n)!}
        &
        n \leq \left\lfloor \frac{|l|}{2} \right\rfloor
        \\
        0 & \text{otherwise}
    \end{cases}
\end{align}
\end{widetext}
which reproduce equivalent expressions as those in~\cite{linton2010}.

\begin{table*}
    \caption{Possible simplifications in the case $\bm{r}_\perp = 0$ for different lattices}
    \label{tab:simplification}
    \begin{ruledtabular}
        \begin{tabular}{cccl}
            Space dim. $d$ & Lattice dim. $d'$ & Lattice position & Simplification \\
            \hline
            3 & 2 & $z = 0$
            & \makecell{
                $Y_{lm}(\theta_{-\bm{r}_\parallel - \bm{R}}, \phi_{-\bm{r}_\parallel - \bm{R}})
                = Y_{lm}\left(\frac{\pi}{2}, \phi_{-\bm{r}_\parallel - \bm{R}}\right)$ \\
                $= \begin{cases}
                    \frac{\sqrt{\frac{2l+1}{4\pi} (l - m)! (l + m)!} (-1)^\frac{l + m}{2}}{2^l \left(\frac{l + m}{2}\right)! \left(\frac{l - m}{2}\right)!}
                    \mathrm{e}^{\mathrm{i} m \phi_{-\bm{r}_\parallel - \bm{R}}} & l + m~\text{even} \\
                    0 & l + m~\text{odd}
                \end{cases}$
            }
            \\
            3 & 1 & $x = 0 = y$
            & $Y_{lm}(\theta_{-\bm{r}_\parallel - \bm{R}}, \phi_{-\bm{r}_\parallel - \bm{R}})
                = \sqrt{\frac{2l+1}{4\pi}} (\sgn((-\bm{r}_\parallel - \bm{R})\bm{\hat{z}}))^l$
            \\
            2 & 1 & $y = 0$
            & $\mathrm{e}^{\mathrm{i} m \phi_{-\bm{r}_\parallel - \bm{R}}} = (\sgn((-\bm{r}_\parallel - \bm{R})\bm{\hat{x}}))^l$
        \end{tabular}
    \end{ruledtabular}
\end{table*}

Due to the properties of the spherical harmonics and the complex exponential function, there can also be some simplifications for the real space sum if there is no perpendicular component of the shift $\bm r_\perp = 0$, and the lattice is placed along certain high symmetry directions, which is done for the derivation of the reciprocal space integral anyhow. These simplifications are listed in \cref{tab:simplification}.

\section{Direct computation with averaging over oscillations}\label{app:convolution}
\begin{figure*}
\includegraphics[width=\linewidth]{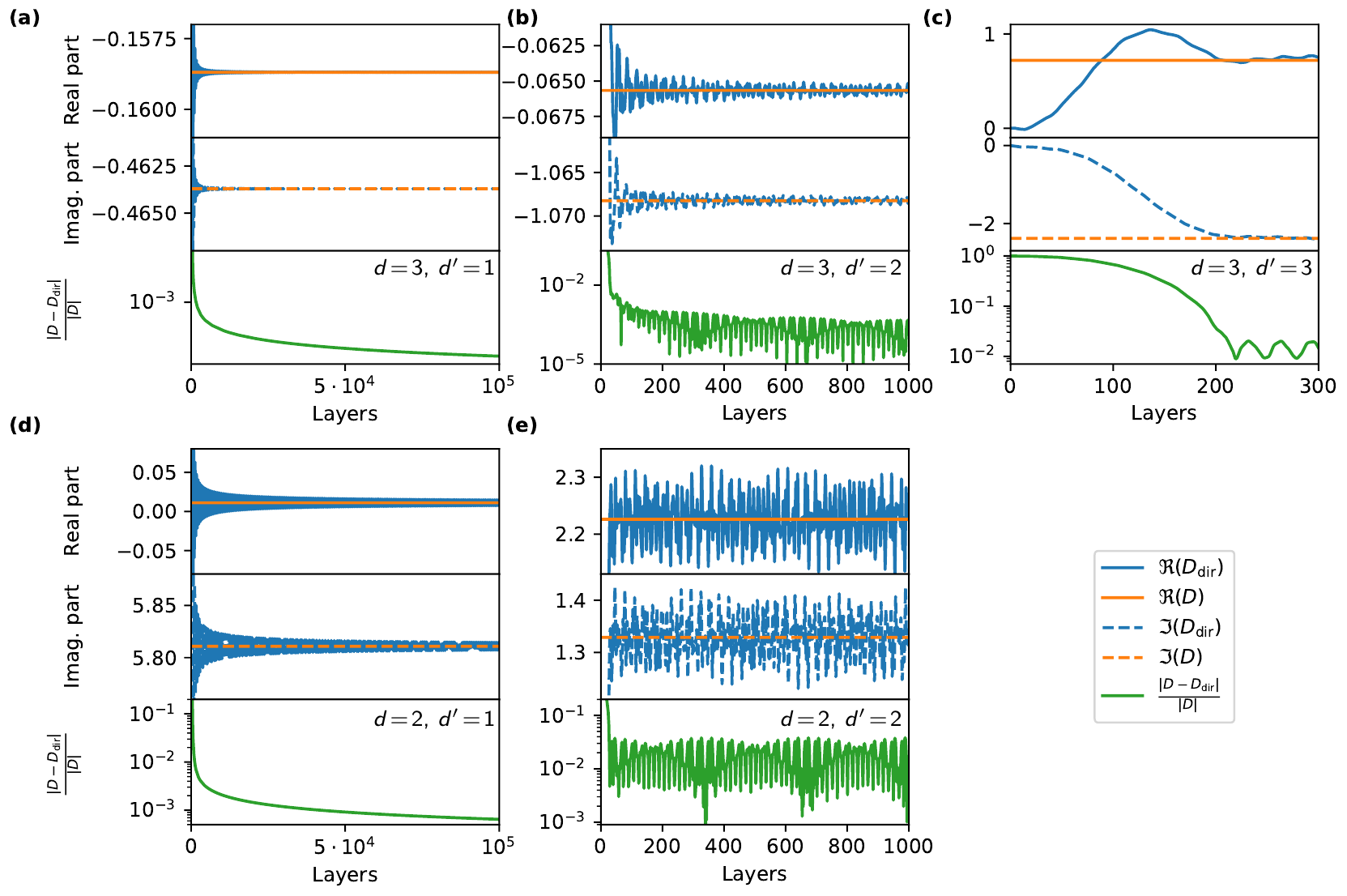}
\caption{The same parameters as in \cref{fig:all} are used but here we use a convolution to average over the oscillations to obtain a faster convergence of the direct summation. Still, the direct approach needs significantly longer.}\label{fig:convolution}
\end{figure*}

To improve the convergence of the direct summation, it is possible to average over oscillations. The results shown in \cref{fig:convolution} are obtained from the data in \cref{fig:all}.

\end{document}